\documentclass[twocolumn]{aastex62}
\usepackage{amsfonts,amsmath,amssymb,amsthm,epsfig,graphicx,float,tabularx}
\begin{document}

\title{A diversity of wave-driven pre-supernova outbursts}

\correspondingauthor{Samantha Wu}
\email{scwu@astro.caltech.edu }

\author{Samantha Wu}
\affiliation{California Institute of Technology, Astronomy Department, Pasadena, CA 91125, USA}

\author{Jim Fuller}
\affiliation{TAPIR, Walter Burke Institute for Theoretical Physics, Mailcode 350-17, California Institute of Technology, Pasadena, CA 91125, USA}

\begin{abstract}
    Many core-collapse supernova progenitors show indications of enhanced pre-supernova (SN) mass loss and outbursts, some of which could be powered by wave energy transport within the progenitor star. Depending on the star's structure, convectively excited waves driven by late stage nuclear burning can carry substantial energy from the core to the envelope, where the wave energy is dissipated as heat. We examine the process of wave energy transport in single-star SNe progenitors with masses between $11 \text{--} 50 \, M_{\odot}$. Using MESA stellar evolution simulations, we evolve stars until core collapse and calculate the wave power produced and transmitted to the stars' envelopes. These models improve upon prior efforts by incorporating a more realistic wave spectrum and non-linear damping effects, reducing our wave heating estimates by $\sim$ 1 order of magnitude compared to prior work. We find that waves excited during oxygen/neon burning typically transmit $\sim \! 10^{46\text{--}47}$ erg of energy at $0.1-10$ years before core collapse in typical ($M < 30 \, M_\odot$) SN progenitors. High-mass progenitors can often transmit $\sim \! 10^{47\text{--}48}$ erg of energy during oxygen/neon burning, but this tends to occur later, at about $0.01-0.1$ years before core collapse. Pre-SN outbursts may be most pronounced in low-mass SN progenitors ($M \lesssim 12 \, M_\odot$) undergoing semi-degenerate neon ignition, and in high-mass progenitors ($M \gtrsim 30 \, M_\odot$) exhibiting convective shell mergers. 
\end{abstract}

\section{Introduction}

Evidence continues to mount that a substantial fraction of core-collapse supernovae (SNe) are preceded by greatly elevated mass loss rates from their progenitor stars. In most cases, this is inferred from SN observations that reveal large amounts of circumstellar material (CSM) very close (within $\sim \! 10^{15}$ cm) to the progenitor star, which is not predicted by standard stellar and wind models. The CSM is usually manifested by faster rise times and brighter early-time SN light curves, or by blue and featureless early spectra indicative of a shock heated envelope or CSM. Narrow flash-ionized emission lines are sometimes seen in the early spectra and then disappear, and they are thought to be produced by confined CSM which is then swept up by the SN ejecta.

Recent examples of SNe with evidence for early interaction include type II SNe such as SN2016bkv \citep{Nakaoka2018}, SN2018zd \citep{Zhang2020a}, and many others \citep{Khazov2016,Forster2018}; and type I SNe such as LSQ13abf \citep{Stritzinger2019}, LSQ13ddu \citep{Clark2020}, SN2018bgv and SN2018don \citep{Lunnan2019}, SN2018gep \citep{Ho2019}, SN2019dge \citep{Yao2020}, SN2019uo \citep{Gangopadhyay2020}. These SNe extend across virtually all spectroscopic classes of SNe (including type II-P, II-L, IIn, Ib, Ic, Ibn, Ic-BL, superluminous Ic, etc.). There are many other SNe that show narrow emission lines and late-time interaction, such as SN2004dk \citep{Mauerhan2018,Pooley2019}, SN 2010bt \citep{Elias-Rosa2018}, SN2012ab \citep{Bilinski2018}, SNhunt151 \citep{Elias-Rosa2018a}, SN2013L \citep{Taddia2020a}, ASASSN-15no \citep{Benetti2018}, iPTF16eh \citep{Lunnan2018}, SN2017dio \citep{Kuncarayakti2018}, and SN2017ens \citep{Chen2018}, indicative of extreme pre-SN mass loss occurring $\sim$decades before the SN.

The list above includes only events from the last three years, and many others are listed in \citet{fuller2017} and \citet{fuller2018}. In several SNe, pre-SN outbursts have been observed directly, and \cite{Ofek2014} find that the majority of type IIn SNe exhibit bright ($L \gtrsim 3 \times 10^7 \, L_\odot$) outbursts in the final months of their lives.

However, it is also important to note that many (perhaps the majority) of SN progenitors do not exhibit pre-SN outbursts or amplified pre-SN variability. \cite{Samson2017} found no significant variability of the progenitor of type Ic SN2012fh, and \cite{Johnson2018}  found variability amplitudes less than $\sim$10\% for four type II-P SN progenitors. \cite{Kochanek2016a} examined some of the best available progenitor constraints for several well studied nearby SNe, finding no evidence for outbursts. 

One should also note that for type II SN with early peaks in their light curves, the CSM is not necessarily related to elevated levels of mass loss, but could instead be produced by moderate amounts of mass ($M \lesssim 1 \, M_\odot$) in an optically thin stellar chromosphere or corona \citep{Dessart2017,Hillier2019} that is not included in standard stellar models.

The inevitable conclusion reached from these observations is that the massive star progenitors of SNe are diverse, with some presenting bright pre-SN outbursts, others exhibiting elevated levels of mass loss, and many more demonstrating no unusual behavior at all. A successful physical explanation for elevated pre-SN mass loss must account for this diversity.

One compelling model for at least some pre-SN outbursts is the wave heating model of \cite{quataert2012}. Vigorous core convection (often carrying more than $10^9 \, L_\odot$) excites internal gravity waves (IGW) that couple with acoustic waves to  deposit a small fraction of this power in the outer layers of the star, which can be sufficient to eject mass or drive an observable outburst. \cite{shiode2014} examined approximate wave heating energetics and time scales in a suite of models and found larger amounts of wave heating in more massive stars, whose outbursts occur closer to core-collapse. 

\cite{fuller2017}  examined the consequences of wave heating in a $15 \, M_\odot$ red supergiant star, finding that waves can inflate the envelopes and possibly drive mass loss through a secondary shock. In compact hydrogen-poor stars, \cite{fuller2018} found that wave heat can launch a dense super-Eddington wind carrying more than $10^{-2} \, M_\odot$/yr. Both studies predicted large changes in luminosity and temperature during phases of enhanced mass loss, although the outburst luminosities in those works did not reach the level of $\sim \! 3 \times 10^7 L_\odot$ seen in type IIn progenitors \citep{Ofek2014}.

However, the amount of wave heat deposited in the envelope is sensitive to the wave spectrum excited by convection, to non-linear wave breaking effects, and to the rapidly evolving core structure. Most prior work has not simultaneously accounted for all of these effects to predict the wave heating rate as a function of time. In this paper, we improve upon prior work through a more complete modeling of the physics at play, applying these calculations to a suite of stars extending over the mass range $M_{\rm ZAMS} = 11-50\,  M_{\odot}$. We find that wave heating rates have been overestimated in some prior work and are not high enough to produce large outbursts in most stars. Interestingly, however, we find that wave-driven outbursts are likely to be most energetic and most prevalent in the lowest and highest-mass SN progenitors.

\section{Implementation of Wave Physics in Stellar Models}

\subsection{Wave Generation and Propagation}
\label{sec:relevantequations}
To implement wave energy transport, we follow the same basic procedure as \citet{fuller2017} and \citet{fuller2018}, which is largely based on the initial work of \citet{quataert2012} and \citet{shiode2014}. We calculate the wave heating rates, but do not simulate the impact of wave heating on the stellar structure.

\begin{table}
\begin{tabular}{cccc}
\hline
Model & 12 $M_{\odot}$ & 40 $M_{\odot}$ & \\
$L_{\rm con}$ & $\sim 5\times 10^9 L_{\odot}$ & $\sim 10^{10} L_{\odot}$ &\\
$\mathcal{M}_{\rm con}$ &  $\sim 0.03$ & $\sim 0.04$ & \\
$f_{\rm esc, \ell=1}$ & $\sim 0.7$ &  $\sim 0.7$ &\\
$f_{\rm esc, \ell=2}$ & $\sim 0.2$ & $\sim 0.1$ & \\
$f_{\rm esc, \ell=3}$ & $\sim 0.05$ & $\sim 0.01$ &\\
Accumulated $E_{\rm heat}$  & $\sim 10^{47}$ erg & $\sim 2\times 10^{47}$ erg & \\
$t_{\rm prop, g}$ &  hours & hours & \\
Global $t_{\rm dyn}$ & months & months &\\
$t_{\rm burn, O/Ne}$ &$\sim 6$ yr &  months & \\
$t_{\rm burn, Si}$ &  weeks & days & \\
$t_{\rm prop, SN} $ & $\sim 15$ hr & $\sim 15$ hr &\\ 
\hline\\[-1mm]
\end{tabular}
\caption{Order-of-magnitude values of some relevant properties and timescales that describe the wave heating phenomenon in two fiducial models, a low-mass $12\, M_{\odot}$ and a high-mass $40\, M_{\odot}$ model. For wave heating due to O/Ne burning phases, we compare the convective luminosity, convective Mach number, and escape fraction for angular wavenumbers $\ell = 1$, $2$, and $3$ (Equation \ref{eq:heatfrac}). We also list the accumulated wave energy at $\sim 1$ day before core collapse (Equation \ref{eq:lheat}), the propagation timescale for gravity waves, the global dynamical time for a red supergiant, the timescales of O/Ne and Si burning, and the timescale for SN shock propagation. }
\label{tab:heatingvals}
\end{table}

To provide a qualitative description of the wave heating phenomenon, we list some order-of-magnitude estimates of relevant quantities and timescales for two fiducial models, a low-mass $12\, M_{\odot}$ model and a high-mass $40\, M_{\odot}$ model in Table \ref{tab:heatingvals}. When appropriate, quantities are estimated during oxygen/neon (O/Ne) burning phases, which we ultimately find to contribute most significantly to wave heating.

As we explain in more detail below, the waves will initially propagate as gravity waves through the core on a timescale $t_{\rm prop, g}\sim r/v_g$ (Equation \ref{eq:gravityvg}). In both models, the waves have plenty of time to escape the core because $t_{\rm prop, g} \ll t_{\rm burn, O/Ne}$ and $t_{\rm prop, g}\ll t_{\rm burn, Si}$, where $t_{\rm burn, O/Ne}$ and $t_{\rm burn, Si}$ are defined as the time until core collapse when O/Ne and Si ignite. Once waves reach the envelope, they propagate as acoustic waves on approximately the global dynamical timescale of the star $t_{\rm dyn} \sim (G\rho)^{-1/2}$. Since $t_{\rm burn, O/Ne},\, t_{\rm burn, Si} < t_{\rm dyn}$ in the high-mass model, an outburst in high-mass supergiants is unlikely. However, in a stripped SN progenitor, the absence of envelope shortens $t_{\rm dyn}$ to minutes, making outbursts much more promising. We have also provided the timescale for SN shock propagation after core collapse for a red supergiant, $t_{\rm prop, SN} \sim \! R/v_{\rm ej}$, assuming $v_{\rm ej} \sim 10^9$ cm/s; it is small compared to the timescale of acoustic wave propagation in our supergiant models.

Gravity waves are excited at the interface between convective and radiative zones and carry a fraction of the kinetic energy of turbulent convection. While convective wave excitation is not totally understood, it is generally agreed that the power put into waves, $L_{\rm wave}$, is at least 
\begin{equation}
\label{eq:lwave}
    L_{\rm wave} = \mathcal{M}_{\rm con} L_{\rm con}
\end{equation}
where $\mathcal{M}_{\rm con}$ is the MLT convective mach number and $L_{\rm con}$ is the convective luminosity \citep{Goldreich1990}. In fact, \citep{Lecoanet2013} predict a somewhat higher flux, depending on the details of the radiative convective interface. We define the convective velocity  $v_{\rm con}$ as
\begin{equation}
\label{eq:Lcon}
    v_{\rm con} = \left(\frac{ L_{\rm con}}{4\pi\rho r^2}\right)^{1/3} \, ,
\end{equation}
and we define the associated MLT convective turnover frequency as
\begin{equation}
\label{eq:omega}
    \omega_{\rm con} = 2\pi \frac{v_{\rm con}}{2 \alpha_{\rm MLT} H},
\end{equation}
where $\alpha_{\rm MLT} H$ is the mixing length and $H$ is the scale height.

At the excitation region, energy is supplied to gravity waves in a power spectrum over frequency and angular wavenumber $\ell$, whose details remain poorly understood. We adopt the spectrum from \citep{Goldreich1990,Shiode2013}:
\begin{align}
\label{eq:ellpowerspectrum}
    \frac{d\dot{E}_g}{d \ln \omega d \ln \ell} &\sim  
    L_{\rm wave} \left( \frac{\omega}{\omega_{\rm{con}}}\right)^{\!-a}
    \left( \frac{\ell}{\ell_{\rm con}} \right)^{\!b+1}
    \left(1+\frac{\ell}{\ell_{\rm con}}\right) \nonumber \\
    & \times \exp \left[-\left(\frac{\ell}{\ell_{\rm con}}\right)^{\!2} \left(\frac{\omega}{\omega_{\rm con}}\right)^{\!-3} \right] \, .
\end{align}
Here, $\ell_{\rm con} = r/H$ is evaluated at the edge of the convective zone and the predicted exponents are $a=13/2$ and $b=2$. We simplify the calculation by setting $\omega = \omega_{\rm con}$ for all $\ell$ values.

In case the scale height $H$ is larger than the size of the convective zone $\Delta r$, we take $\ell_{\rm con} = r/\min(\Delta r,H)$.
The proportion of wave energy generation per $\ell$ value, $\dot{E}_\ell/L_{\rm wave}$, can be calculated using this power spectrum, which we normalize to Equation \ref{eq:lwave}, i.e., we set $\sum \dot{E}_\ell = L_{\rm wave}$.

\begin{figure}
    \centering
    \includegraphics[width=\columnwidth]{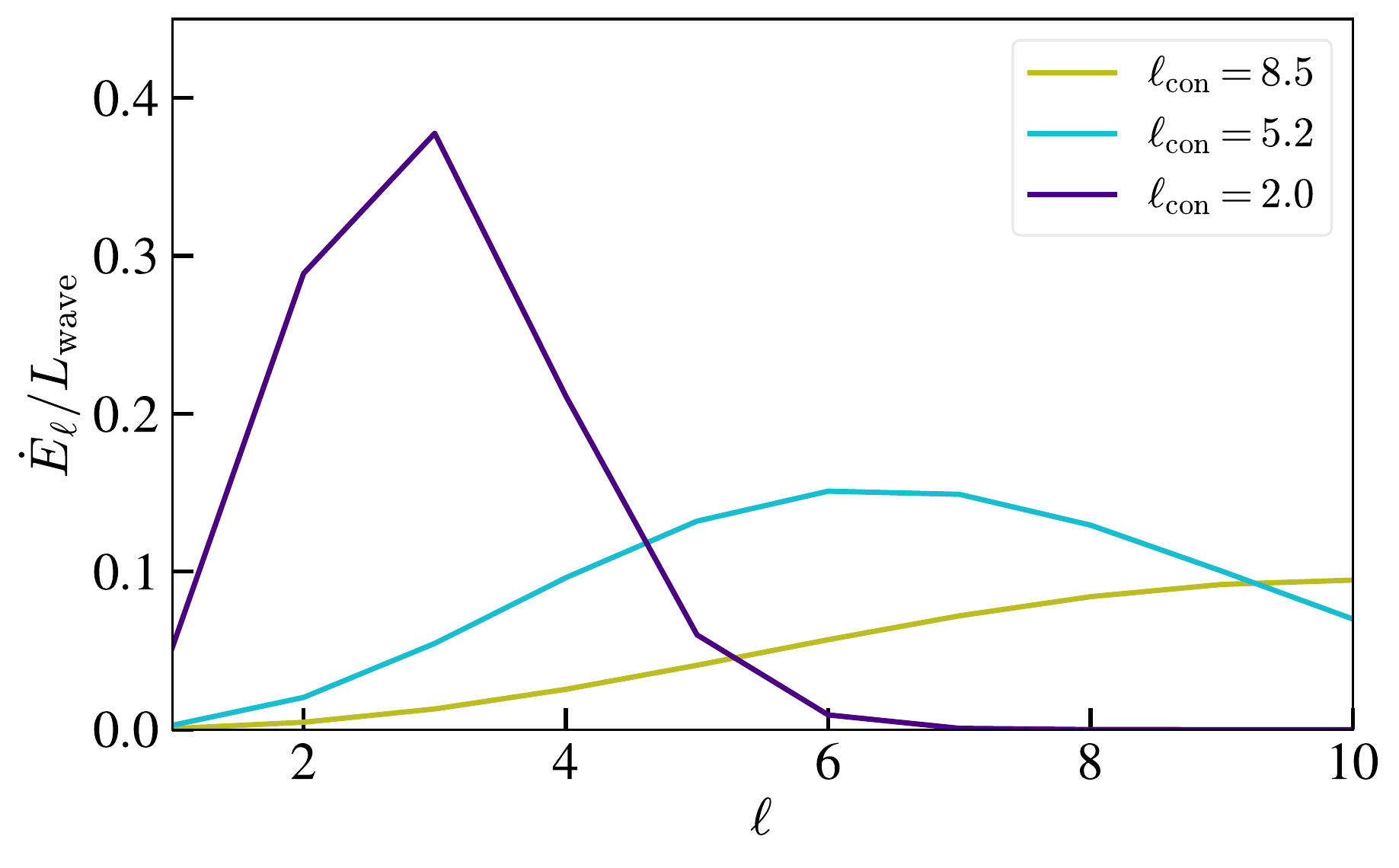}
    \caption{ Fraction of wave energy generation per $\ell$ value $\dot{E}_\ell/L_{\rm wave}$, shown for representative values of $\ell_{\rm con}= r/\min(\Delta r,H)$. Large values of $\ell_{\rm con}$ represent thin convective shells and vice versa.
    We find $\ell_{\rm con} \sim 2\text{--}4$ for core O/Ne and Si burning, while He shell burning in our higher mass models can exhibit values as high as $\ell_{\rm con} \approx 8$.
    }
    \label{fig:edotell}
\end{figure}

In the WKB limit, waves that remain linear have the dispersion relation
\begin{equation}
\label{eq:dispersion}
    k_r^2=\frac{(N^2-\omega_{\rm wave}^2)(L_{\ell}^2-\omega_{\rm wave}^2)}{\omega_{\rm wave}^2 c_s^2},
\end{equation}
where $N^2$ is the squared Brunt-V\"ais\"al\"a frequency, $k_r$ is the radial wavenumber, and $L_{\ell}^2 = {\ell}({\ell}+1)c_s^2/r^2$ is the Lamb frequency squared. In the limit that $\omega_{\rm wave} \ll N, L_{\ell}$, this reduces to the gravity wave dispersion relation
\begin{equation}
\label{eq:gravitykr}
    k_r^2=\frac{\ell(\ell+1)N^2}{\omega_{\rm wave}^2 r^2}
\end{equation}
with group velocity 
\begin{equation}
\label{eq:gravityvg}
    v_g=\omega_{\rm wave}^2 r/\sqrt{\ell(\ell+1)N^2}.
\end{equation}
The limit $\omega_{\rm wave} \gg N, L_{\ell}$ gives acoustic waves, with dispersion relation
\begin{equation}
    k_r^2=\frac{\omega_{\rm wave}^2}{c_s^2}
\end{equation}
and group velocity $v_g=c_s$. In either of these limits, linear waves propagate freely and approximately conserve their luminosity, apart from damping effects discussed below.

If $\omega_{\rm wave} > N$ and $\omega_{\rm wave} < L_{\ell}$ or vice versa, then the radial wavenumber is imaginary and waves are evanescent. The probability of tunneling through this evanescent zone, or the fraction of transmitted wave energy, is approximately given by the transmission coefficient
\begin{equation}
\label{eq:transmissioncoeff}
    T^2 = \exp\left(-2 \int_{r_0}^{r_1}  \lvert k_r \rvert dr \right),
\end{equation}
where the integral is taken over the evanescent zone. In practice, waves sometimes encounter multiple evanescent zones and the thickest evanescent zone dominates the wave reflection  (see Appendix B2 of \citealt{fuller2017}). To calculate the wave flux tunneling into the envelope, we thus take the minimum value of $T^2$ out of all evanescent zones (i.e., the thickest evanescent region), $T_{\rm min}^2$.  The remaining fraction $1-T^2$ of wave energy that encounters the evanescent region is reflected from the boundary of the evanescent zone.

\begin{figure}
    \centering
    \includegraphics[width=\columnwidth]{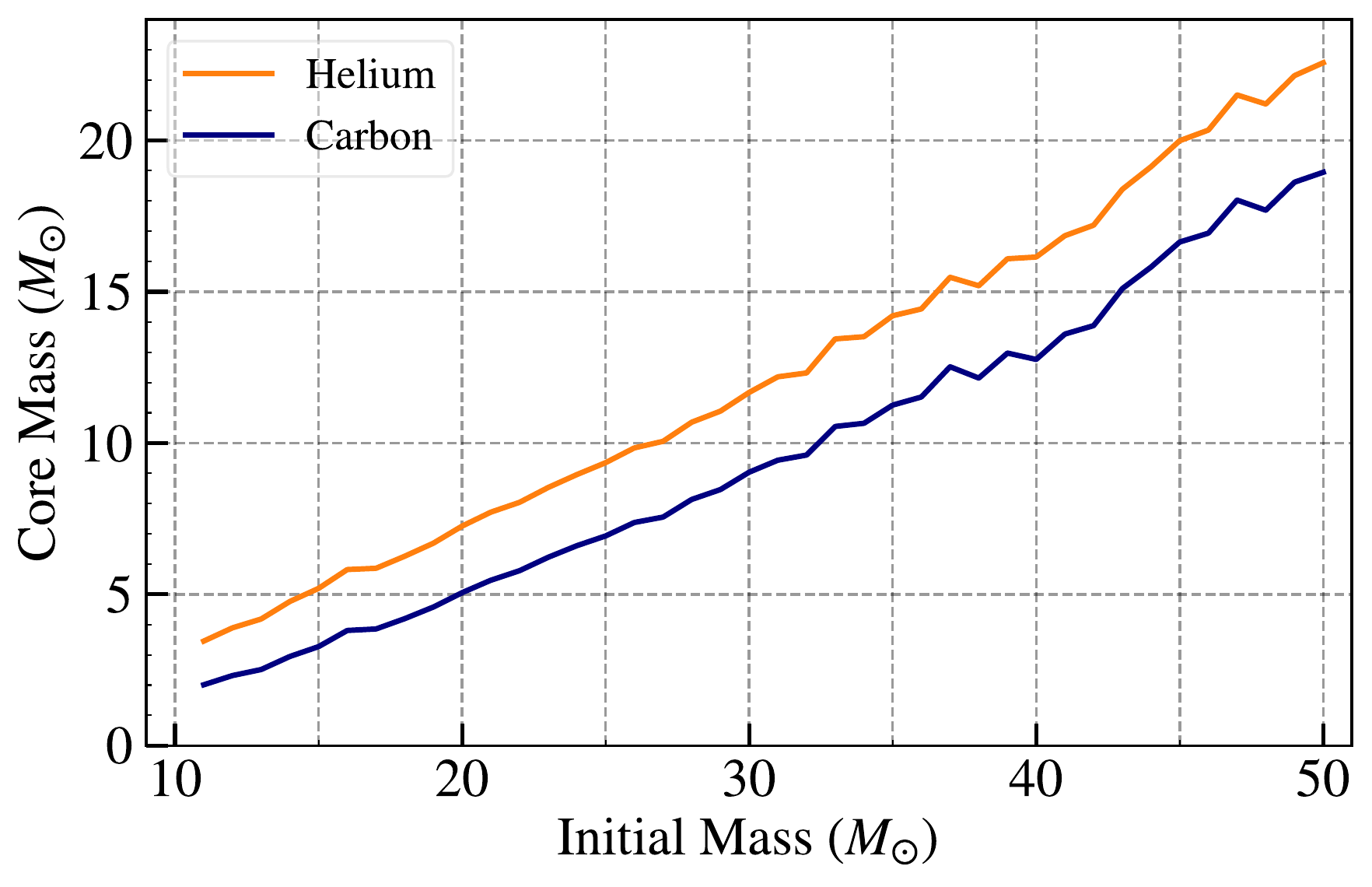}
    \caption{The helium (He) core mass (orange) and carbon (C) core mass (blue) as a function of initial progenitor mass for our models.  }
    \label{fig:corevsinitial}
\end{figure}

\begin{figure*}
    \includegraphics[width=\textwidth]{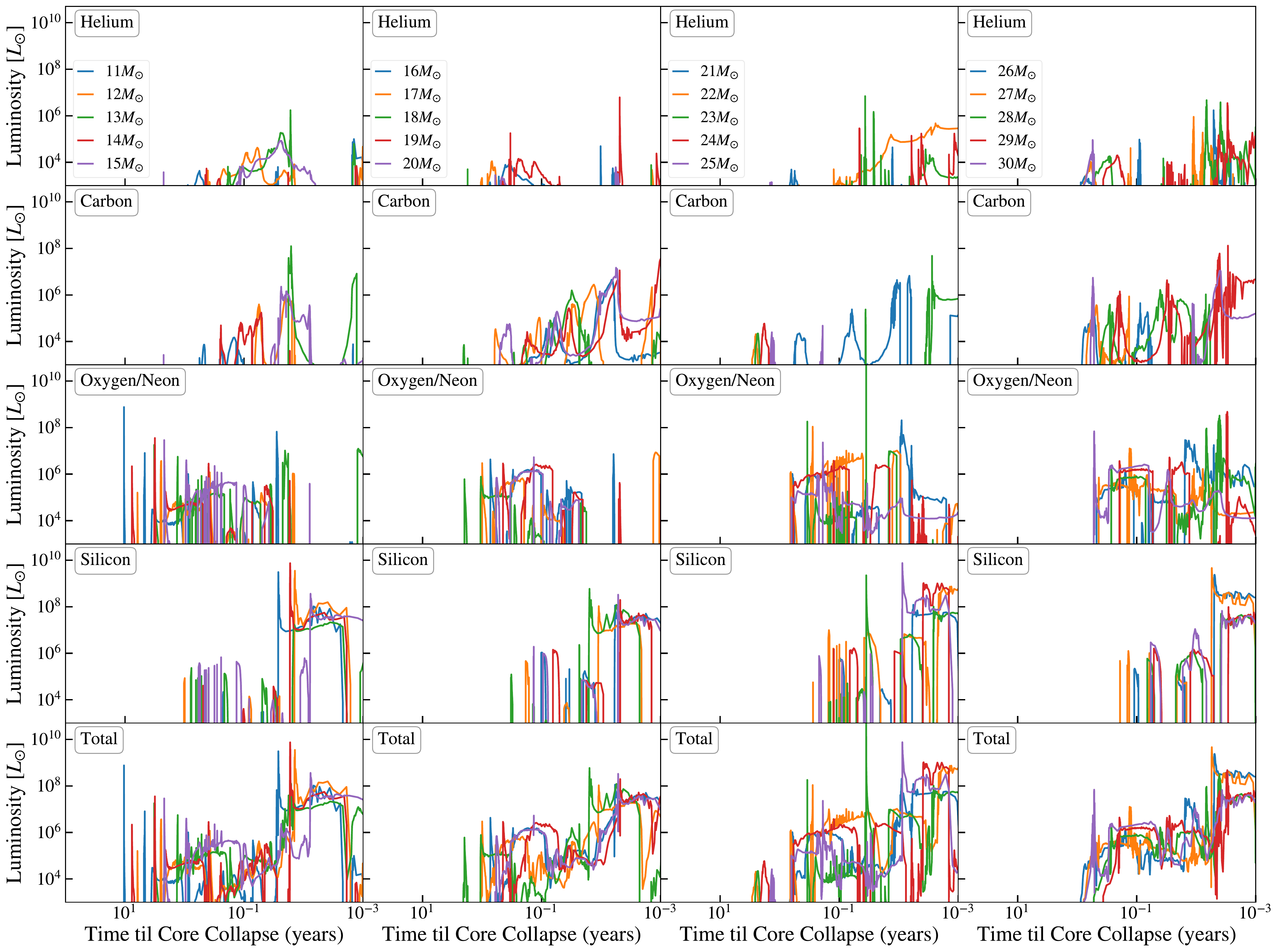}
    \caption{Wave heating luminosity for models with initial mass in the range $11\text{--}30\,  M_{\odot}$, grouped into columns by mass. Element labels in each row refer to the type of burning occurring in the convective region that generates the waves. Sharp spikes in wave heating luminosity typically occur at the ignition of a new burning phase or during a convective shell merger (Section \ref{sec:notablemodels}).}
    \label{fig:11to30Lwaves}
\end{figure*}

\begin{figure*}
    \centering
    \includegraphics[width=\textwidth]{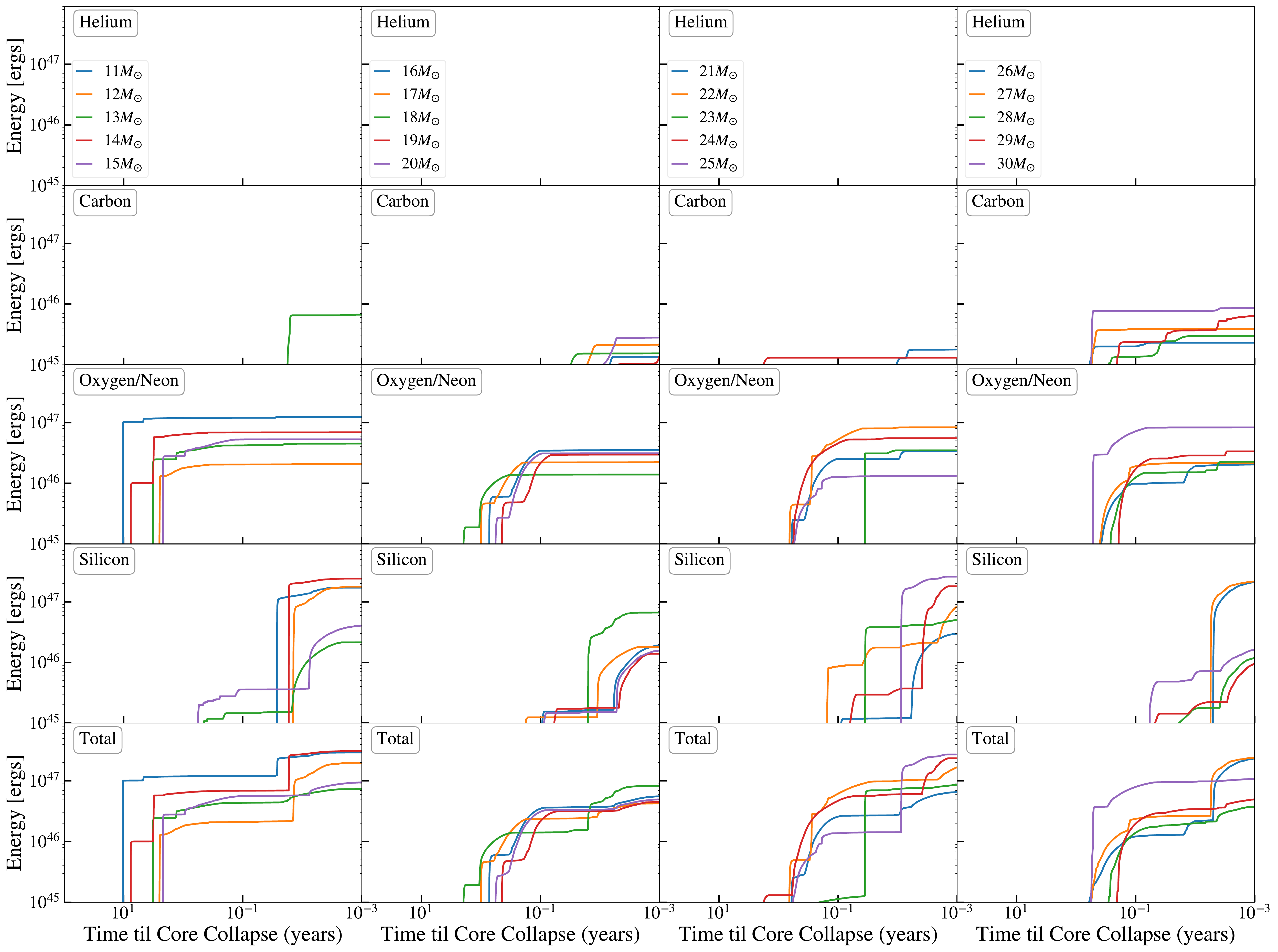}
    \caption{Accumulated wave heating energy transported to the envelope, for models corresponding to the heating rates of Figure \ref{fig:11to30Lwaves}. }
    \label{fig:11to30Ewavesreduced}
\end{figure*}

The fraction of wave energy that escapes  the core, $f_{\rm esc}$, is determined by the transmission coefficient (Equation \ref{eq:transmissioncoeff}) and energy losses within the core. Neutrino emission attenuates the net energy flux escaping from the g-mode cavity, and the local neutrino wave damping rate is given by
\begin{equation}
\label{eq:nudamprate}
    \gamma_{\nu} = \frac{\Gamma_1^2 \nabla_{\rm ad}^2 g^2 }{N^2 c_s^4} \left( \frac{\partial \ln \epsilon_{\nu}}{\partial \ln T} \right)_{\rho} \epsilon_\nu,
\end{equation}
where all quantities are as defined in Appendix B of \citet{fuller2017}.
Gravity waves are also damped by the diffusion of photons, and the thermal damping rate is given by
\begin{equation}
\label{eq:photdamprate}
    \gamma_{\rm rad} \simeq k_r^2 K,
\end{equation}
where 
\begin{equation}
\label{eq:thermaldiff}
    K =  \frac{16 \sigma_{SB} T^3}{3 \rho^2 c_p \kappa}
\end{equation}
is the thermal diffusivity.

As a result, after traversing to the upper edge of the core and back, a wave's energy is attenuated by the factor
\begin{equation}
\label{eq:fnu}
    f_{\nu} = e^{x_{\nu}} = \exp\left[2\int_{r_{-}}^{r_{+}} \frac{\gamma_{\nu} + \gamma_{\rm rad}}{v_g} dr\right]
\end{equation}
where $v_g$ is the gravity wave group velocity (Equation \ref{eq:gravityvg}) and the integral is taken over the upper and lower boundaries of the gravity wave cavity. 
Then the escape fraction is 
\begin{equation}
\label{eq:heatfrac}
    f_{\rm esc, \ell} = \left(1+\frac{f_{\nu}-1}{T_{\rm min}^2} \right)^{-1}.
\end{equation}
and the $\ell$-dependent power that escapes to heat the envelope is
\begin{equation}
    L_{\rm heat, \ell} = f_{\rm esc, \ell} \dot{E}_\ell.
\end{equation}

Another source of energy loss is non-linear wave breaking. To calculate the gravity wave non-linearity in the WKB limit, we first calculate the radial Lagrangian displacement $\xi_r$. Assuming a constant wave energy escape rate, then the rate at which energy enters the cavity $L_{\rm heat, \ell}$ equals the rate $\dot{E}_{\rm leak}$ at which the energy  $E$ in the g-mode cavity leaks out:
\begin{equation}
    \dot{E}_{\rm leak} = \frac{E}{t_{\rm leak}}.
\end{equation}
As in \citet{fuller2015}, the energy leakage timescale is
\begin{equation}
    t_{\rm leak} =\frac{ 2 t_{\rm cross}}{T^2},
\end{equation}
and the energy per unit radius 
\begin{equation}
    E_r = \frac{E}{v_g t_{\rm cross}}
\end{equation}
is given by $E_r = 4\pi r^2 \rho \omega^2 \xi^2$ in the cavity, where $\xi$ is the wave displacement amplitude.
Using the fact that the radial displacement $\lvert \xi_r \rvert \simeq \frac{\omega}{N} \lvert \xi \rvert$ for gravity waves, we can rearrange to find 
\begin{equation}
\label{eq:displacement}
    \left|\xi_r\right|^2 = \frac{2}{T_{\rm min}^2}\frac{L_{\rm heat, \ell}}{4 \pi r^2\rho v_g N^2} \, .
\end{equation}
Then a measure of the gravity wave non-linearity as a function of $\ell$ is 
\begin{equation}
\label{eq:nonlinearity}
    \left|k_r \xi_r\right| = \left[\frac{2}{T_{\rm min}^2}\frac{L_{\rm heat, \ell} N [\ell (\ell+1)]^{3/2}}{4\pi\rho r^5\omega^4} \right]^{1/2}
\end{equation}

Where $\left|k_r \xi_r\right| \geq 1$, waves are highly non-linear, whereas linear waves have $\left|k_r \xi_r\right| \leq 1$. Highly non-linear gravity waves will break and their energy will cascade to small scales, where the energy dissipates and thermalizes on a wave crossing time scale \citep{Barker2010}. This process caps wave amplitudes at $\left|k_r \xi_r\right| \sim 1$. Waves that require large amplitudes $\xi_r$ to sustain their power and frequency are potentially non-linear. Since non-linear terms couple waves of different $\ell$, it is not clear what the appropriate non-linear breaking threshold is for a spectrum of waves, but if a wave of any $\ell$ value has $\left|k_r \xi_r\right| \geq 1$, it is likely that a non-linear cascade will damp the energy of all the waves on a short time scale. We account for this non-linear damping by capping the wave amplitude as described below.
\begin{figure*}
    \includegraphics[width=\textwidth]{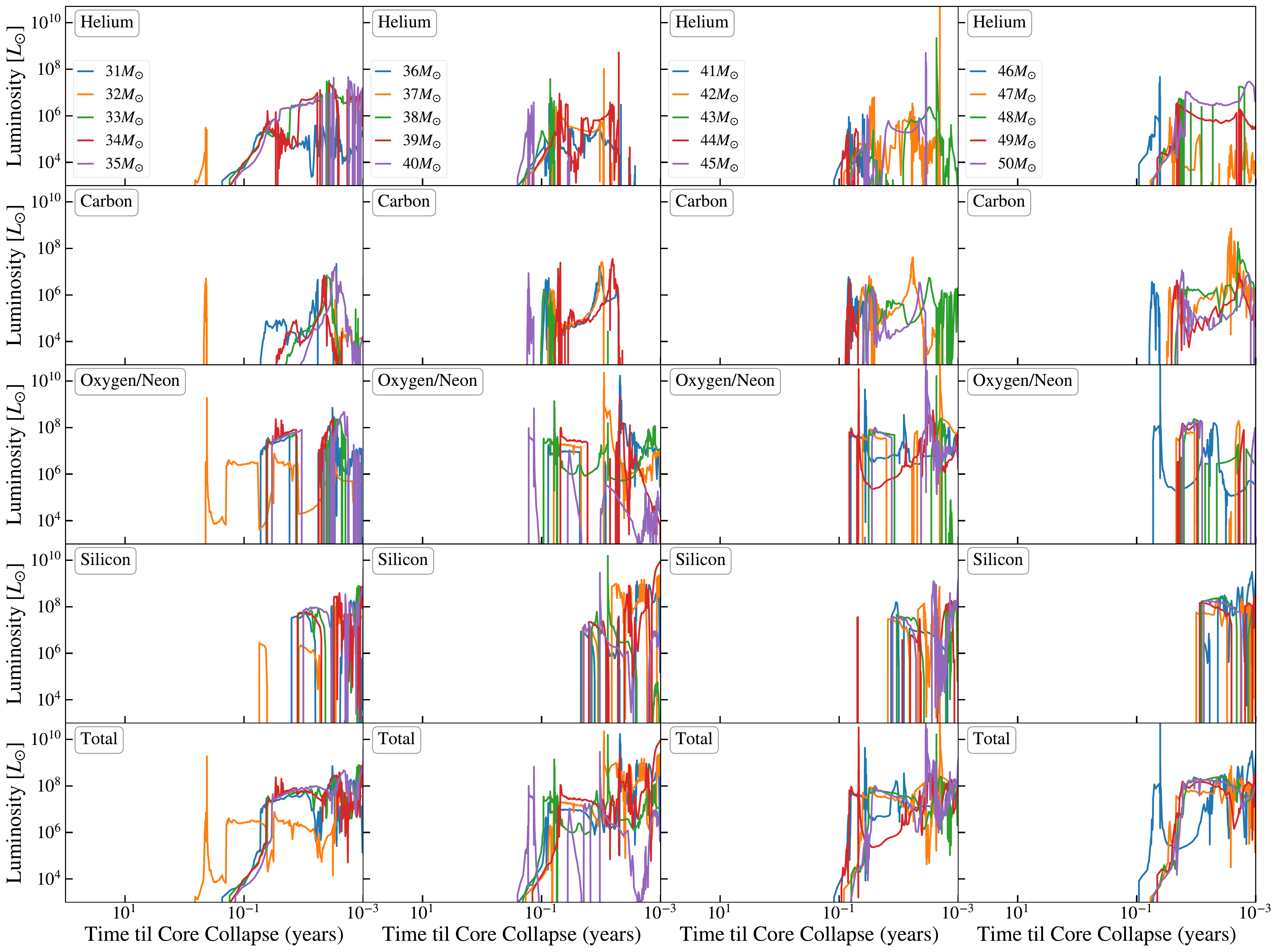}
    \caption{Same as Figure \ref{fig:11to30Lwaves}, but showing the wave heating luminosity for models with initial mass in the range $31\text{--}50\,  M_{\odot}$.}
    \label{fig:31to50Lwaves}
\end{figure*}
\begin{figure*}
    \centering
    \includegraphics[width=\textwidth]{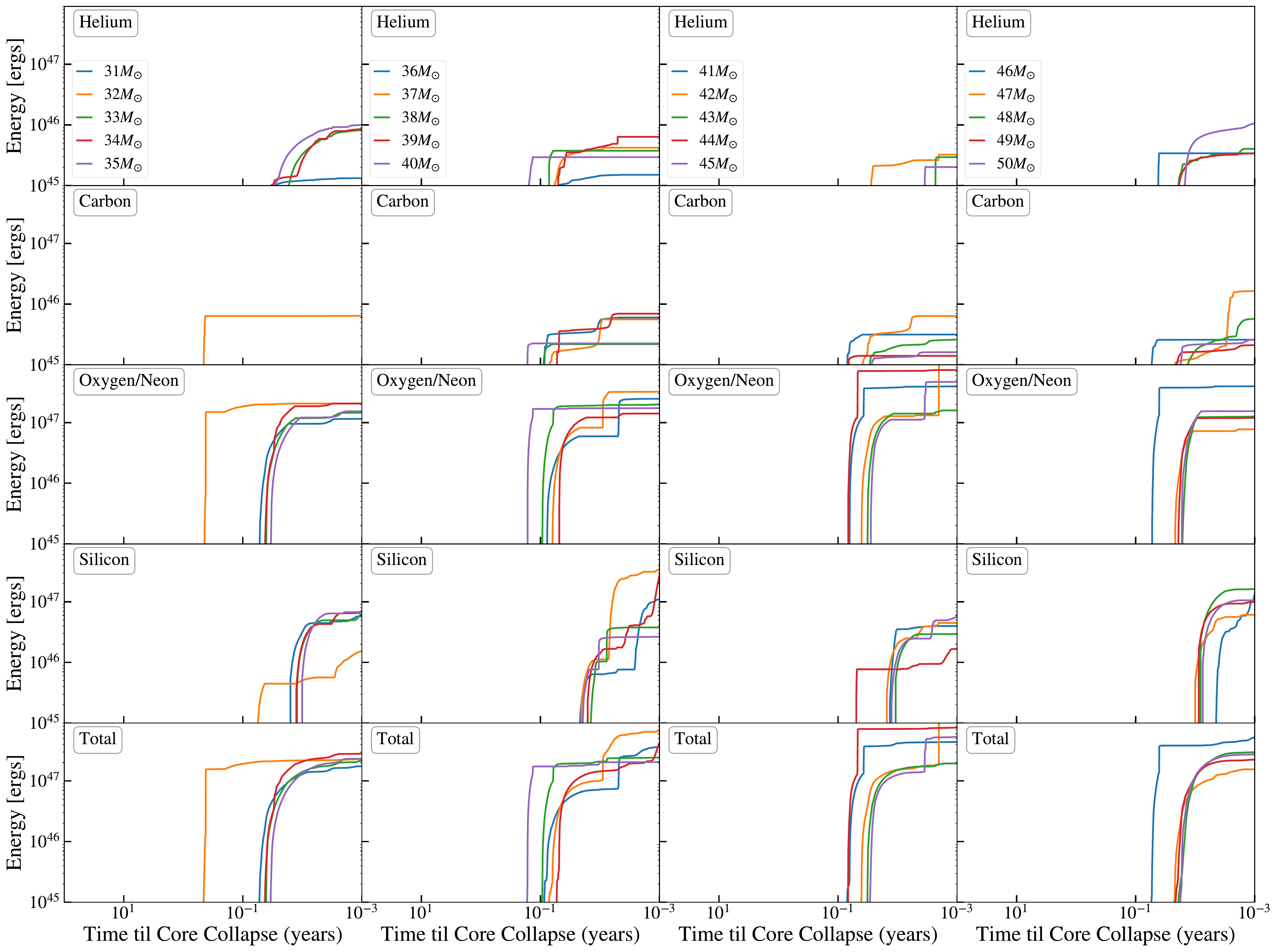}
    \caption{Same as Figure \ref{fig:11to30Ewavesreduced}, but for the accumulated energy of higher mass models in the range $31\text{--}50\,  M_{\odot}$.}
    \label{fig:31to50Ewavesreduced}
\end{figure*}
\begin{figure*}
    \centering
    \includegraphics[width=0.492\textwidth]{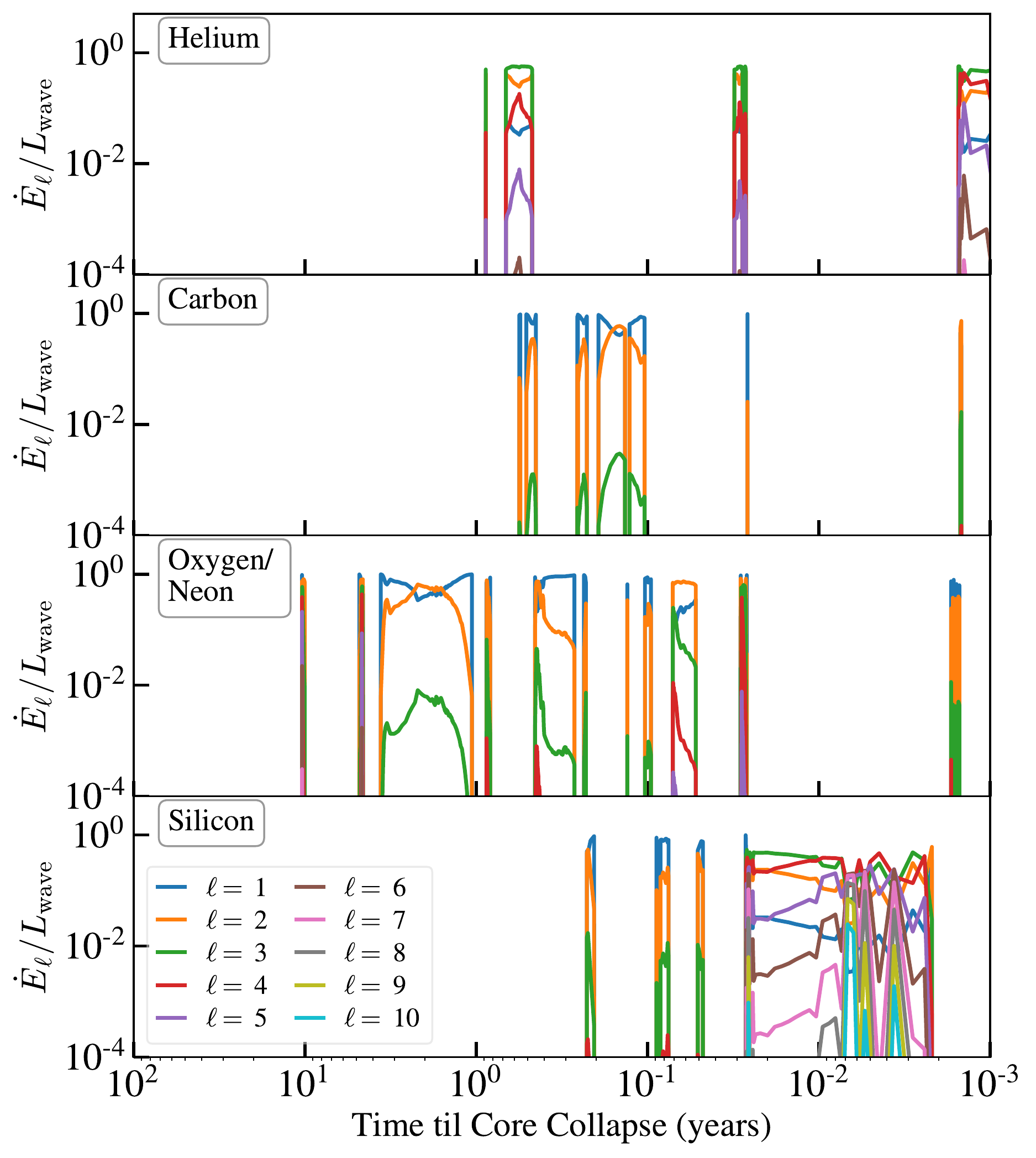}
    \includegraphics[width=0.49\textwidth]{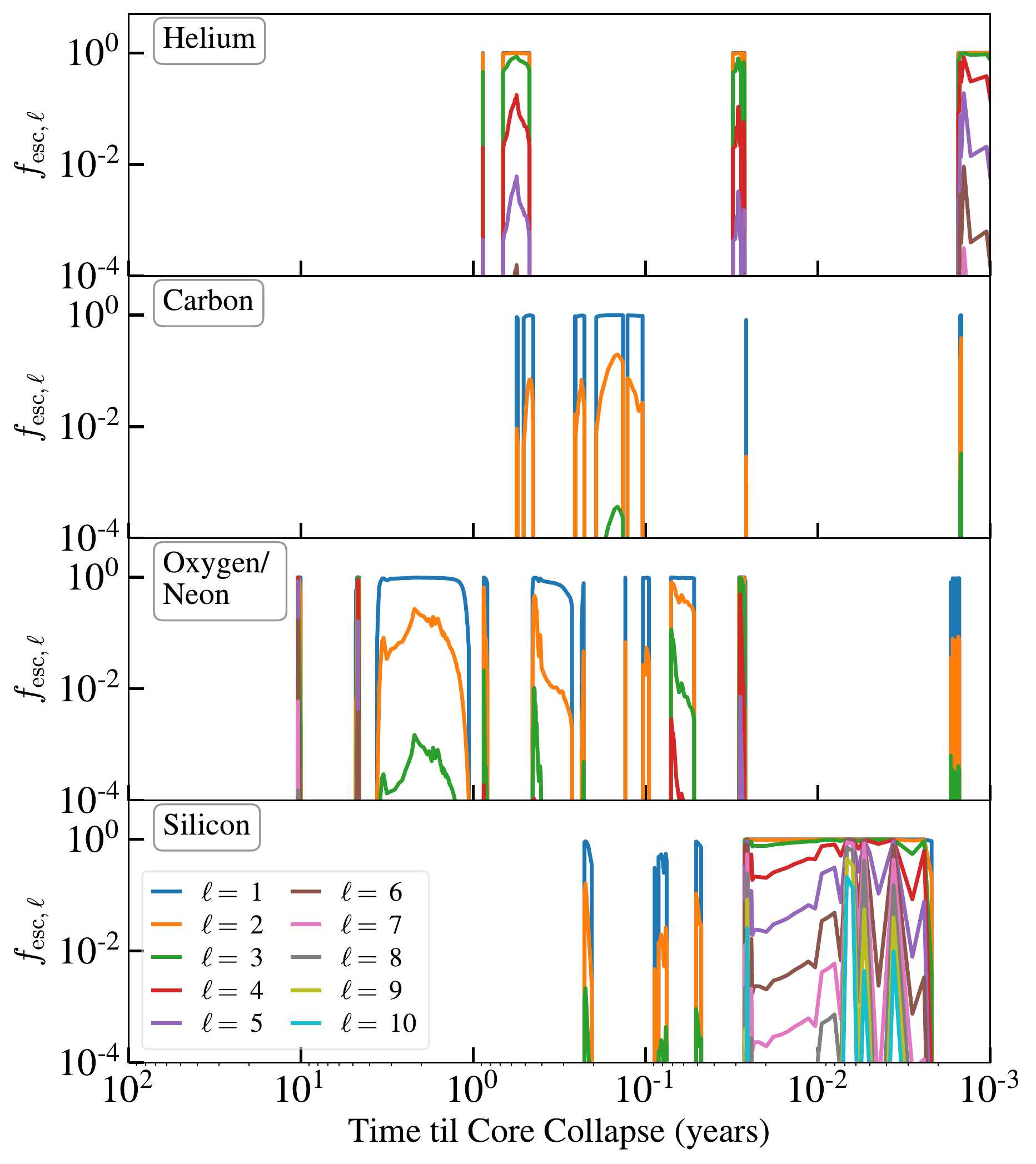}
    \caption{Fraction of total wave heat deposited in the envelope by waves of each angular number $\ell$ (left), and the wave escape probability for each $\ell$ (right), for the $11\,  M_{\odot}$ model shown in Figures \ref{fig:11to30Lwaves} and \ref{fig:11to30Ewavesreduced}. Results are only shown for waves where $L_{\rm heat,\ell} > 10^3\, L_{\odot}$.}
    \label{fig:11fesc}
\end{figure*}

\subsection{Methods}
\label{sec:methods}
We run a suite of MESA simulations \citep{mesa2011,mesa2013,mesa2015,mesa2018,mesa2019} for zero-age main sequence (ZAMS) masses between 11 and 50 $M_{\odot}$ and evolve the stars from the main sequence to core collapse.\footnote{Model parameters are available at https://zenodo.org/communities/mesa.}
The evolution of late burning stages in the core is primarily determined by the helium (He) core mass and the carbon (C) core mass, each shown in Figure \ref{fig:corevsinitial} as a function of initial (ZAMS) mass. Here, the He core mass is the mass coordinate where the He mass fraction is $\geq 0.01$ and the hydrogen mass fraction has dropped below $0.01$; similarly, the C core mass defines the transition where the He mass fraction $< 0.01$ and C mass fraction $\geq 0.01$. 

At each timestep, we perform the following calculations for the convective burning regions in each model.

\begin{enumerate}
    \item Calculate wave luminosity and frequency: We find the mass-weighted average of the total wave luminosity and convective frequency within each convective burning region (Equations \ref{eq:lwave} and \ref{eq:omega}). We assume here that the waves of each $\ell$ value are of the same frequency $\omega = \omega_{\rm con}$, and we consider $\ell = 1$ to $10$ waves.
    \item Calculate transmission coefficient and neutrino damping factor for each $\ell$ value: For each $\ell$ value, we integrate over the evanescent regions above each convective burning region, using Equation \ref{eq:transmissioncoeff} to find the transmission coefficient $T^2$ through each evanescent region. As explained in Section \ref{sec:relevantequations}, we use the minimum $T_{\rm min}^2$ when there are multiple evanescent regions. For the wave energy attenuated due to local neutrino damping and radiative diffusion damping, the damping rates due to these effects are given by Equations \ref{eq:nudamprate} and \ref{eq:photdamprate}. We integrate equation \ref{eq:fnu} through all overlying gravity wave cavities and calculate the attenuation factor $f_{\nu}$ for each $\ell$ value.
    \item Calculate wave heating rate and energy transmitted: Given the transmission coefficient $T^2$ and the neutrino damping factor $f_{\nu}$, the fraction of wave luminosity that can escape is given by Equation \ref{eq:heatfrac} for each $\ell$ value. The total heating luminosity $L_{\rm heat}$ for each convective burning region is calculated by summing up the wave energy generation rate per $\ell$ value multiplied by the escape fraction per $\ell$ value, i.e. $L_{\rm heat} = \sum_{\ell = 1}^{10} f_{\rm esc, \ell} \dot{E}_{\ell} $. Then the energy transmitted to the envelope at each time step is 
    \begin{equation}
        \label{eq:lheat}
       \Delta E_{\rm heat} = L_{\rm heat} dt,
    \end{equation}
    evaluated for each convective burning region that generates waves. 
    \item Calculate wave non-linearity: Section \ref{sec:relevantequations} introduces $|k_r\xi_r|$ as a measure of the nonlinearity of the gravity waves and notes that waves of different $\ell$ are coupled by nonlinear effects. As a result, we consider the largest value of $|k_r\xi_r|$ out of all the waves of different $\ell$. If this maximum $|k_r\xi_r|>1$ in the gravity wave cavity, the wave amplitudes are likely capped such that $|k_r\xi_r| \lesssim 1$. The wave power is proportional to the square of the wave amplitude, so in the case of non-linear waves we reduce $L_{\rm heat}$ by a factor of $|k_r\xi_r|^2$. While we are not able to capture the complexities of how non-linear coupling among waves of different $\ell$ truly affects wave heating, this approach at least provides us with an understanding of where non-linear effects are most important.
    
\end{enumerate}

\section{Results}
\label{sec:results}

Figure \ref{fig:11to30Lwaves} shows the wave heating rate $L_{\rm heat}$ (Equation \ref{eq:lheat}) of models in the mass range $11\text{--}30\, M_{\odot}$, and Figure \ref{fig:11to30Ewavesreduced} shows the cumulative energy transmitted by waves to the envelope, $\int L_{\rm heat} dt$, for this mass range. Each row corresponds to the wave heat generated by different convective burning regions, He, C, O/Ne, and Si, throughout the stars' lifetimes. In our models, the largest power is usually produced by O/Ne and Si burning regions, whereas the typical power from He and C burning is $1\text{--}2$ orders of magnitude lower. 

The mass range $11\text{--}20\, M_{\odot}$ typically generates $10^{5\text{--}6}\, L_{\odot}$ of wave power from O/Ne burning between $0.01\text{--}10$ years before core collapse, with brief excursions above $10^{6\text{--}7} \, L_\odot$ that typically occur at the ignition of core or shell-burning phases. In more massive stars, between $21 \text{--}30\, M_{\odot}$, waves from O/Ne burning typically carry $10^{6\text{--}7}\, L_{\odot}$ from $\sim \! 0.1$ years before core collapse onward. The spikes in wave energy generation rates from O/Ne burning months to years before core collapse are responsible for the majority of wave heating in the $11\text{--}15\, M_{\odot}$ models, creating the sudden jumps in accumulated energy in Figure \ref{fig:11to30Ewavesreduced}. These sudden jumps are most pronounced in the lowest mass models because O/Ne burning ignites in semi-degenerate conditions, discussed more in Section \ref{sec:notablemodels}. For some models, certain C shell-burning spikes in wave power can be of similar magnitude, but they occur too briefly and too late in the star's lifetime to contribute appreciably to wave heating. Waves from Si burning can carry $1\text{--}2$ orders of magnitude more power than O/Ne burning throughout the $11\text{--}30\, M_{\odot}$ models, but Si burning only sustains this power for days to weeks before core collapse. 

For the high-mass models between $31\text{--}50\, M_{\odot}$, Figure \ref{fig:31to50Lwaves} shows that wave power is generally higher for all burning types, but the high luminosity phases are both brief and late in the star's lifetime. The power from He and C burning waves remains $1\text{--}2$ orders of magnitude lower than that of the other burning regions. The waves produced by O/Ne burning in these high-mass models generate $\sim \! 10^{7\text{--}8}\, L_{\odot}$ of power from a few weeks before core collapse onward. Meanwhile, the power carried by Si burning waves is on the same order as that of O/Ne burning waves, but this power is only sustained for hours to days before core collapse. Throughout the mass range $20\text{--}50\, M_{\odot}$, many models exhibit extremely large spikes in the wave heating rates of O/Ne burning that reach $\sim \! 10^{10}\, L_{\odot}$. One notable example is the spike of the $32\,  M_{\odot}$ model months before core collapse, which is atypically early for its mass range. These spikes are due to convective shell mergers, discussed further in Section \ref{sec:notablemodels}.

The models in the lower mass range $11 \text{--} 30\,  M_{\odot}$ exhibit quite a bit of scatter in the accumulated energy transmitted by waves  (Figure \ref{fig:11to30Ewavesreduced}), as changing the mass by only $1\, M_{\odot}$ alters both the the total amount of energy accumulated and the time before core collapse when the most energy is transmitted. This is particularly evident when looking at the energy transmission from Si burning, which may accumulate $>10^{47}$ erg by $10^{-2}$ years before core collapse in some models, but fails to achieve this in models that only differ by a solar mass. In the upper mass range, Figure \ref{fig:31to50Ewavesreduced} shows that models which are close in mass typically result in similar magnitudes and timescales of energy deposition. However, there is still a fair amount of scatter in the energy scale of Si burning, and there are a few outliers. Compared to the other models in the $31\text{--}35\,  M_{\odot}$ mass range, the $32\,  M_{\odot}$ model accumulates $>10^{47}$ erg of energy earlier in O/Ne burning. Similarly, the $44\,  M_{\odot}$ and $46\,  M_{\odot}$ models have higher and earlier O/Ne heating than neighboring-mass models.

Despite the large scatter, there are general trends within different mass ranges in our models. For the majority of $11\text{--}20\,  M_{\odot}$ models, O/Ne burning dominates the transmitted energy and the energy transmission rises early, at a few months to years before core collapse. The $21\text{--}30\,  M_{\odot}$ models reach similar energy scales to the $11\text{--}20\,  M_{\odot}$ models, but usually this accumulates later at weeks before core collapse. In all models from $11 \text{--} 30\,  M_{\odot}$, there is negligible energy transmission from He burning and very little contribution from C burning as well. 

In the $31\text{--}50\,  M_{\odot}$ range, most models begin to accumulate more energy than the lower-mass models in O/Ne burning. In addition, this energy is transmitted later on average, at several days to weeks before core collapse. He and C burning each still contribute over an order of magnitude less energy in this mass range. Throughout these high-mass models, Si burning consistently accumulates energy in a sharp, late rise at $\sim \! 1 \text{--} 3$ days before core collapse. However, the $\sim \! 10^{47}$ erg of energy accumulated from Si burning is typically a few times less than that deposited by waves from O/Ne burning.
\begin{figure*}
    \centering
    \includegraphics[width=0.49\textwidth]{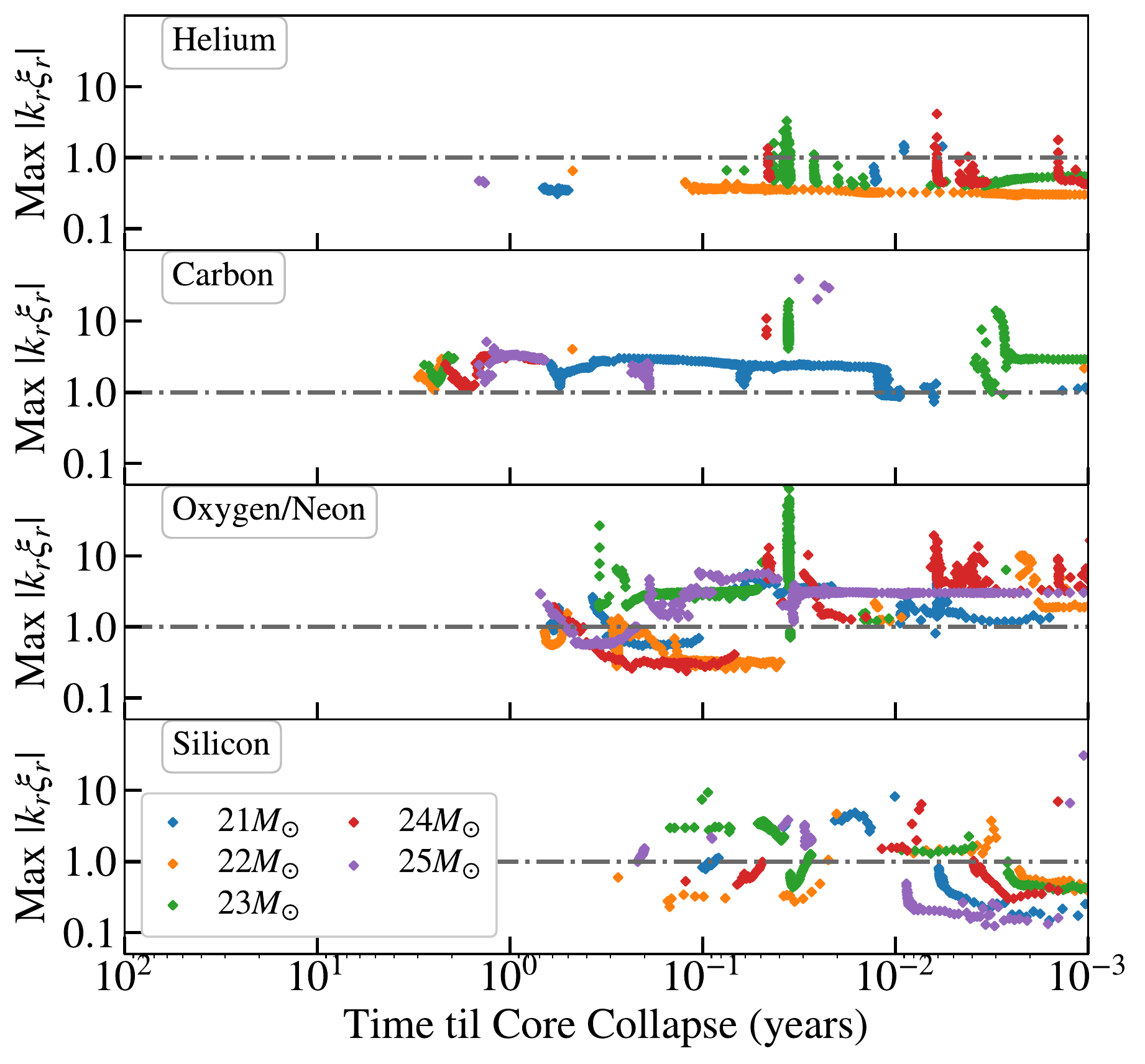}
    \includegraphics[width=0.49\textwidth]{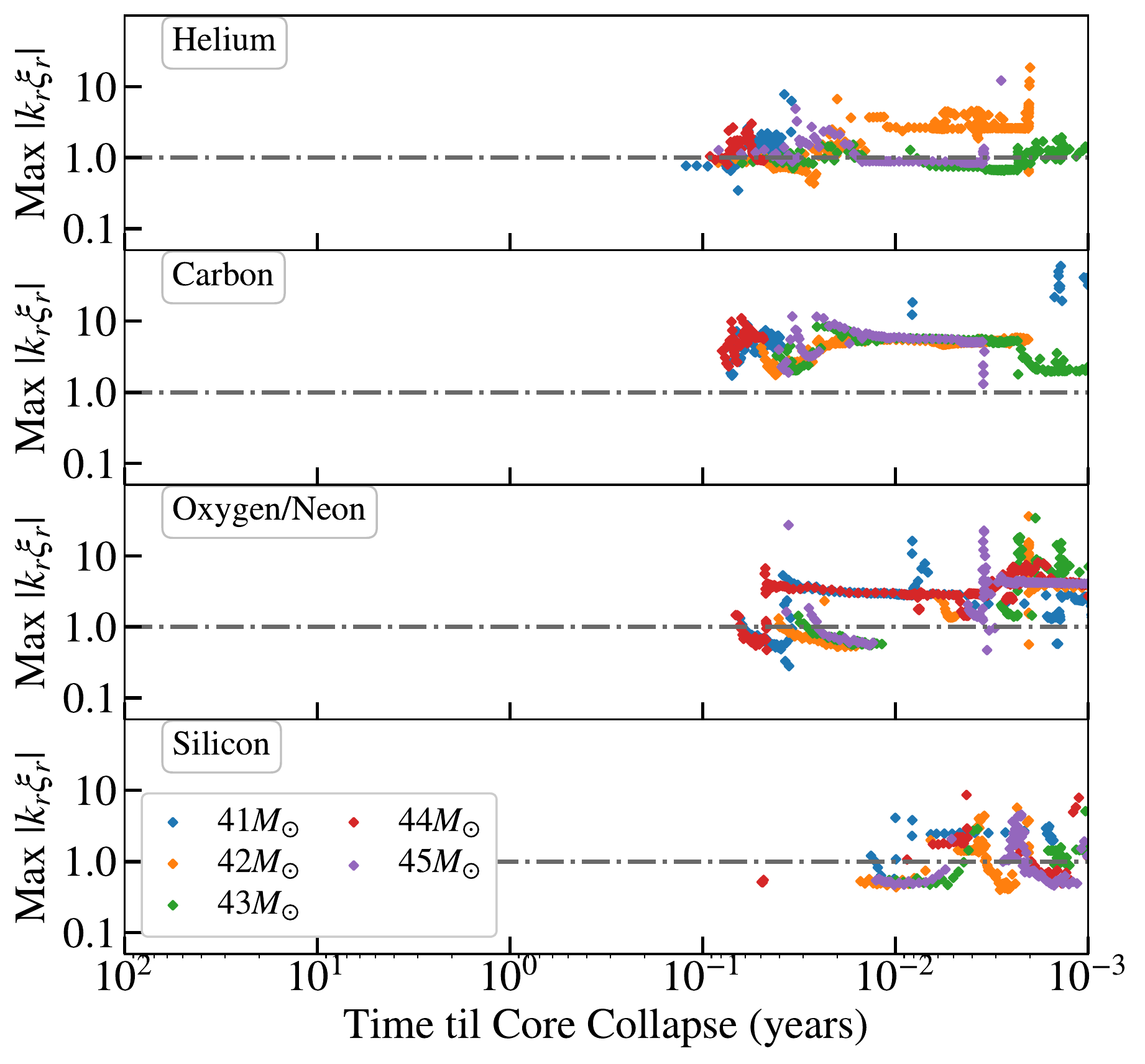}
    \caption{The maximum value of $\left|k_r \xi_r\right|$, taken over waves of all $\ell$, for models with initial mass $21 \text{--} 25\,  M_{\odot}$ (left) and $41 \text{--} 45\,  M_{\odot}$ (right). The gray dashed-dotted line denotes $\left|k_r \xi_r\right| =1$, above which the waves are considered strongly nonlinear so that the wave heating rate is reduced, as described in the text. Results are only shown for waves where $L_{\rm heat,\ell} > 10^3\, L_{\odot}$.}
    \label{fig:krxirs}
\end{figure*}

In most of these models, the heating is dominated by $\ell = 1 \text{--} 3$ waves, as exemplified for the $11 \, M_\odot$ model in Figure \ref{fig:11fesc} (note that the escape and heating fractions are only shown for waves where $L_{\rm heat,\ell} > 10^3\, L_{\odot}$). The right panel shows that low-$\ell$ waves are much more likely to escape before damping, due to their larger transmission coefficients $T^2$ (equation \ref{eq:transmissioncoeff}) and their smaller damping fractions (equation \ref{eq:fnu}). Although the wave power spectrum generated by convection usually peaks for $\ell \gtrsim 3$ (Figure \ref{fig:edotell}) so that more power initially goes into higher $\ell$ waves, the escape fraction is many orders of magnitude smaller for high $\ell$. Consequently, waves of $\ell=1$ and $\ell=2$ (and occasionally $\ell=3$) constitute the large majority of the escaping heat, as shown in the left panel of Figure \ref{fig:11fesc}. Only for helium burning are $\ell = 4,5$ able to contribute comparable amounts. 

Our wave heating rates are about an order of magnitude lower than predicted by previous work. \citet{fuller2017} estimated for a $15\,  M_{\odot}$ red supergiant that waves generated from core O/Ne burning would carry $\sim \! 10^7\, L_{\odot}$ of power and that $\sim 10^{48}$ erg of energy would be deposited into the envelope at months to years before core collapse; in our results, that mass range at best would transmit $10^{47}$ erg on that timescale from O/Ne burning and possibly a few times $10^{47}$ erg much later from Si burning. 
In addition, \citet{shiode2014} found that core O/Ne burning for $12\text{--}30\, M_{\odot}$ models excites waves carrying a few $\times10^{46\text{--}47}$ and few $\times 10^{47}$ erg of energy, and high mass models ($40\, M_{\odot}$ and $50\, M_{\odot}$) carry up to $8\times10^{47}$ erg of wave energy from O/Ne burning. In contrast, we estimate transmitted wave energies that are lower by a factor of a few on average compared to the models in \citet{shiode2014}.

The differences stem from the fact that prior work assumed that most wave power went into $\ell=1$ waves, and previous calculations did not account for non-linear wave dissipation. In our implementation of wave physics, we distribute wave power over a spectrum of different $\ell$-valued waves. Figure \ref{fig:edotell} shows that for typical values of $\ell_{\rm con} \gtrsim 2$, $\ell=1$ waves receive $\lesssim 10 \%$ of the wave flux. This large reduction accounts for most of the differences with prior work, though non-linear wave breaking also plays a role.

\subsection{Non-linearity}

When waves are strongly non-linear such that $\left|k_r \xi_r\right| > 1$ within the core, we reduce the wave heating rate by a factor of $\left|k_r \xi_r\right|^2$ (Section \ref{sec:methods}). As shown in Figure \ref{fig:krxirs}, this reduction can amount to a suppression of $1\text{--}2$ orders of magnitude in some cases. The points in Figure \ref{fig:krxirs} (only plotted for waves where $L_{\rm heat,\ell} > 10^3\, L_{\odot}$ as in Figure \ref{fig:11fesc}) show the maximum value of $\left|k_r \xi_r\right|$ out of waves of all $\ell$ for two mass ranges, $21 \text{--} 25\,  M_{\odot}$ (left) and $41 \text{--} 45\,  M_{\odot}$ (right).
In these mass ranges, as well as for all models with $M > 15\, M_{\odot}$, the waves are usually nonlinear during carbon, oxygen/neon, and silicon burning.
In the upper mass range of $35\text{--}50\,  M_{\odot}$, waves produced by helium burning are also nonlinear throughout the last $0.1\text{--}10$ years until core-collapse.

Reducing the wave power by non-linearity has the most considerable effect on waves generated by convective C shell-burning, since in this case often the waves carrying the most power are also quite non-linear. Before taking non-linearity into account, there were several models in every mass range which could transmit close to $10^{47}$ erg of energy via C shell-burning waves, but non-linear saturation limits the accumulated energy of C burning waves in any model to at most $10^{46}$ erg. The typical value of $\left|k_r \xi_r\right|$ for O/Ne and Si burning waves is similar overall to that of C burning waves, but unlike C burning waves, the O/Ne and Si burning waves that carry the largest $L_{\rm heat, \ell}$ are often less non-linear. Thus in most of the models they are still able to transmit considerable amounts of energy and they usually make the largest contributions to energy transmission in the models. Due to these very energetic waves, the energy transmission from O/Ne and Si burning each remain the same order of magnitude after suppression due to non-linearity in most of the models. Exceptions to this include the $25\, M_{\odot}$ and $27\, M_{\odot}$ models, which experience significant suppression of O/Ne burning due to non-linear effects.
In particular, the O/Ne and Si energy transmission for the $11\text{--}20\,  M_{\odot}$ mass range is not significantly altered by non-linear effects, due to the high wave frequencies in these models.

\begin{figure}
    \centering
    \includegraphics[width=\columnwidth]{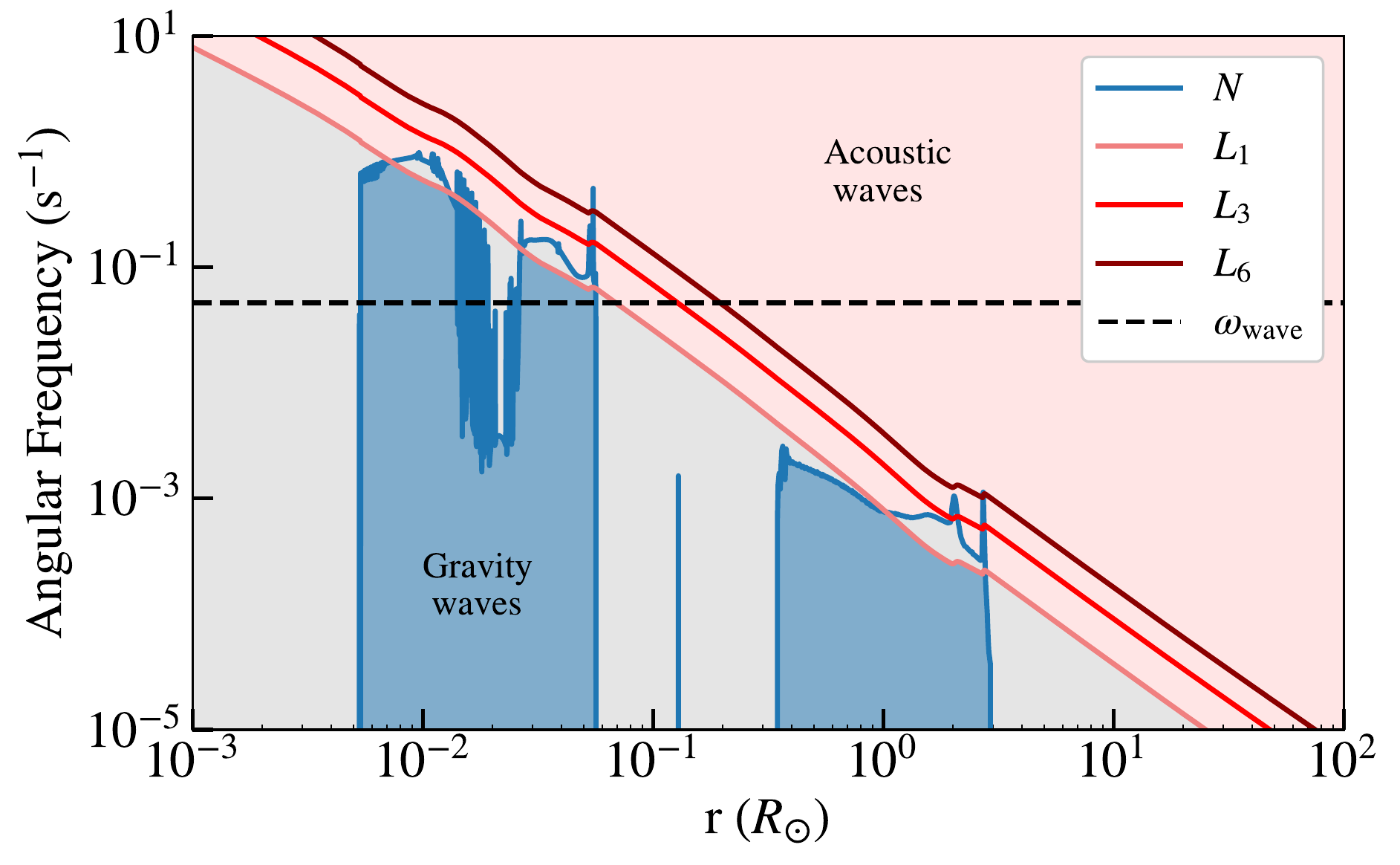}
    \caption{Propagation diagram for the $11\,  M_{\odot}$ model during core Ne burning, showing the Brunt-V\"ais\"al\"a frequency N (blue) and the Lamb frequency $L_{\ell}$ for $\ell = 1,3,$ and $6$ (shades of red). Waves propagate through the gravity wave cavity (blue region) and into the envelope as acoustic waves (red region, shown for $\ell = 1$), tunneling through evanescent zones (gray region) along the way. Semi-degenerate Ne ignition in the core causes vigorous convection, exciting waves with high frequencies of $\omega \sim \! 0.05$ rad/s (dashed line). The high frequencies of these waves allow them to more easily tunnel through a thinner evanescent zone.
    }
    \label{fig:11Mprop}
\end{figure}

\begin{figure}
    \centering
    \includegraphics[width=\columnwidth]{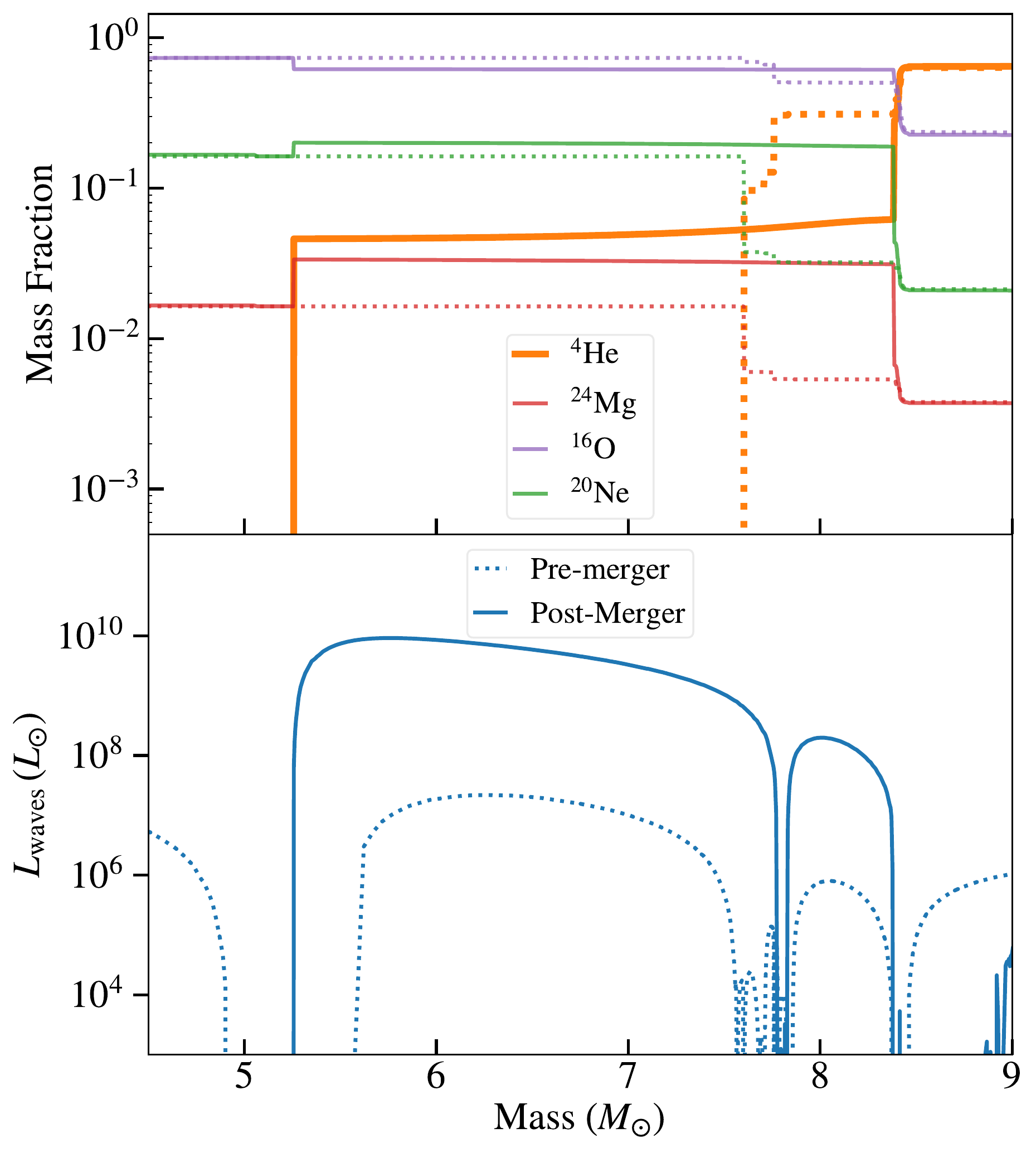}
    \caption{Top: The mass fractions of certain elements near the convective He- and C-burning shells immediately before a shell merger event (dotted lines), and after the shell merger (solid lines). All quantities are shown for a $30\,  M_{\odot}$ model about 2 weeks before core collapse. Note how He is mixed downward into the C-burning shell. Bottom: The wave luminosity (Equation \ref{eq:lwave}) produced by the same convective shell before the shell merger (dotted) and after the shell merger (solid lines). The shell exhibits vigorous burning from $\alpha$-capture reactions post-merger. }
    \label{fig:30Mshellmerge}
\end{figure}

\subsection{Notable Models}
\label{sec:notablemodels}

One of the main new results of our investigation is that the lowest mass models show some of the largest heating rates from O/Ne burning. For example, the 11 $M_{\odot}$ model is noteworthy because its O/Ne heating rate is very large and occurs $\approx \! 10$ years before core collapse, earlier than most models. Figure \ref{fig:11Mprop} shows a propagation diagram for this model, which demonstrates that the frequency of the waves associated with the large O/Ne burning luminosity is high enough to create a thin evanescent region above the core so that waves can easily tunnel into the envelope. As a result, the high wave luminosity is not as greatly reduced by neutrino damping. In addition, the high wave frequency reduces the impact of non-linearity (Equation \ref{eq:nonlinearity}).

The high wave luminosities and frequencies of the $11 \, M_\odot$ model are due to semi-degenerate ignition of Ne in the core; the degeneracy parameter $\eta \sim\! 10$ in the core at Ne ignition. In contrast, for models with $M > 15\,  M_{\odot}$, $\eta \lesssim 1$ or at most $\eta \sim$ a few during core O/Ne burning, so these higher mass models do not exhibit vigorous burning due to semi-degenerate O/Ne ignition. Instances of degenerate Si ignition also lead to very large spikes in wave heating from Si burning for the $11,\, 12$, and $14\,  M_{\odot}$ models (Figure \ref{fig:11to30Lwaves}), and Si ignition in the $20\text{--}25\,  M_{\odot}$ range is also moderately degenerate. However, these energy contributions occur quite late at $\sim\! 0.01$ years before core collapse.

In many of our high-mass models, convective shell mergers occur during the last year before core collapse. Typically this phenomenon occurs between a He and C-burning shell, when He is mixed into the high-temperature C-burning region. This then causes an enormous increase in energy generation due to chains of $\alpha$-capture reactions on C, O, Ne, etc. The heating rates can be anomalously high, e.g., $L_{\rm waves} \gtrsim 10^{9\text{--}10}\,  L_{\odot}$, suddenly increasing the energy transmission for the associated model. 

For example, the jump in C and O/Ne burning for the $30\,  M_{\odot}$ model stems from a shell merger (far right column of Figure \ref{fig:11to30Ewavesreduced}). The top panel of Figure \ref{fig:30Mshellmerge} shows how the shell merger affects the composition of the burning shell. The shell between $5\text{--}8\,  M_{\odot}$ initially contains a negligible amount of helium (dotted yellow line in top panel of Figure \ref{fig:30Mshellmerge}), but the shell merger causes an influx of helium (solid yellow line) and triggers vigorous burning of C, O, Ne, etc. by $\alpha$ capture reactions. Due to the high temperature, these $\alpha$ captures are favored over the typical progression onward from triple-$\alpha$ burning of He. This causes the wave luminosity (Equation \ref{eq:lwave}) generated by the burning shell to jump by 3 orders of magnitude (bottom panel of Figure \ref{fig:30Mshellmerge}). The sudden change in composition and luminosity is representative of the shell merger phenomenon that is seen in many of the higher-mass models. 

These shell mergers are consequential for our models, as the jumps in wave power transmit large amounts of energy that often constitute major contributions to the accumulated energy of the model. In the $21\text{--}30\,  M_{\odot}$ mass range, sudden jumps in O/Ne heating are generally due to shell mergers, with the exception that the jump in energy for 23 $M_{\odot}$ in O/Ne is due to central core ignition after a period of off-center O burning. The $31\text{--}46\,  M_{\odot}$ mass range exhibits shell mergers in all but the $31 \, M_\odot$, $33\text{--}35\,  M_{\odot}$, and $43\, M_{\odot}$ models, and the shell mergers are linked to large spikes in energy in those models.
For $M \geq 47\,  M_{\odot}$, shell mergers occur, but later than $10^{-3}$ years before core collapse. In contrast, no models between $11\text{--}20\,  M_{\odot}$ have shell mergers.

Some of the most extreme jumps in energy caused by shell mergers are ultimately lowered by nonlinear effects so that the accumulated energy transmission becomes more typical of the associated model's mass range. While the vigorous convection excites high-frequency waves that are less prone to nonlinear damping, the enormous wave fluxes do lead to non-linearity. For example, although the energy transmission from the shell merger in the $30\,  M_{\odot}$ model would be unusually large, non-linearity effects limit it to around the typical amounts in this mass range (Figure \ref{fig:11to30Ewavesreduced}). In addition, the $32\,  M_{\odot}$ and $35\text{--}40 \, M_{\odot}$ models would have featured jumps in accumulated energy that were anomalously high without non-linearity reducing them to more typical amounts.

\section{Discussion}
\label{sec:discussion}
\subsection{Outburst Energies and Time Scales}
\label{sec:outbursts}
To assess which models have the greatest potential to produce pre-supernova outbursts from wave heating, we consider the following two quantities: the outburst energy $E_{\rm burst}$,
which we define as the total energy deposited by $10^{-2}$ years before core collapse; and the outburst timescale $t_{\rm burst}$, which we define as the time until core collapse for accumulated wave energy to exceed $10^{47}$ erg. We choose the value $10^{47}$ erg because this is comparable to the amount of energy needed to eject substantial mass from a red supergiant or compact helium star (\citealt{fuller2017,fuller2018}, Linial et al. in prep), and is thus an approximate threshold energy needed to power a pre-SN outburst.

\begin{figure}
    \centering
    \includegraphics[width=\columnwidth]{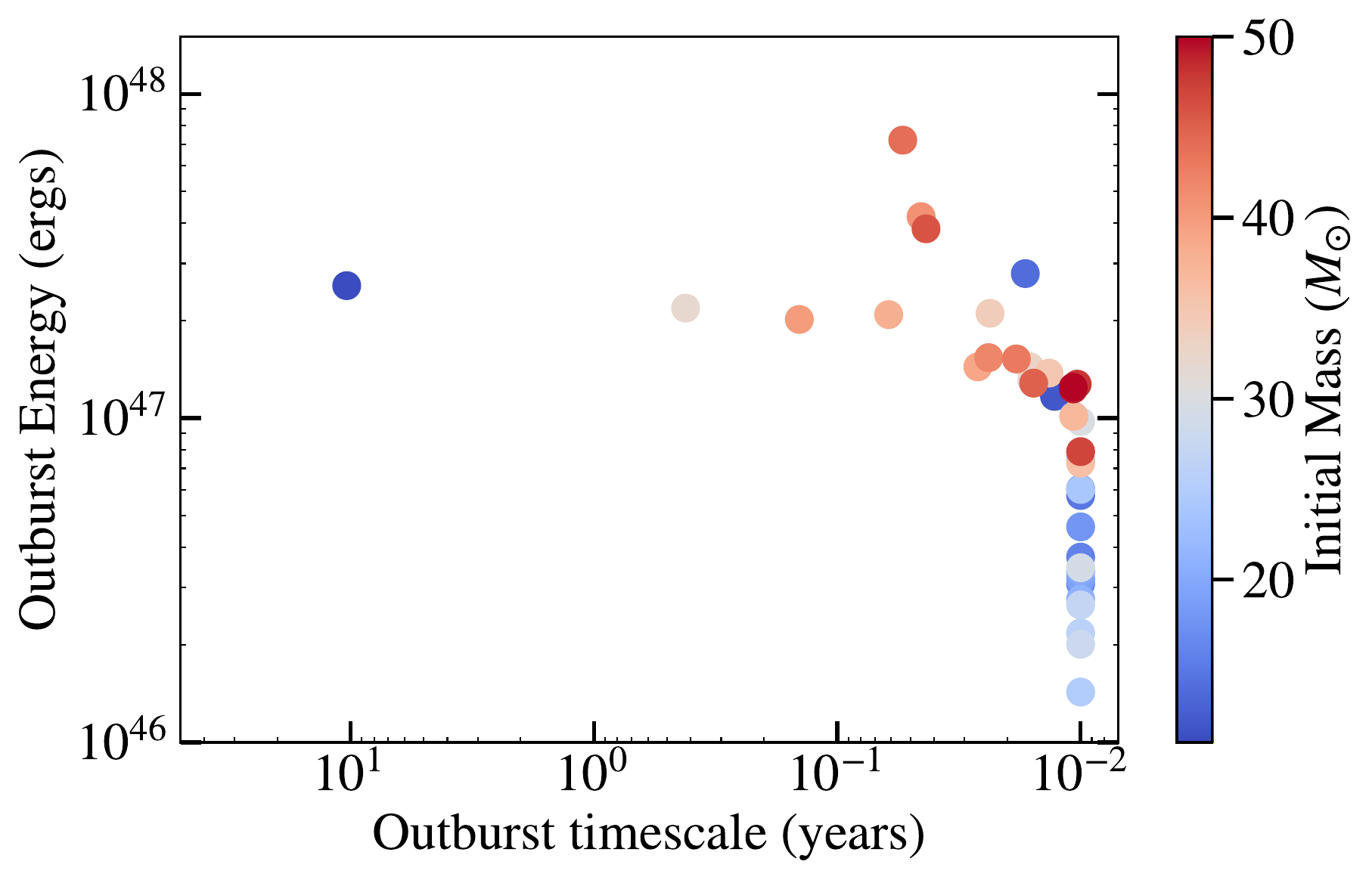}
    \caption{Outburst energy versus outburst timescale for each of our models, as defined in Section \ref{sec:outbursts}. The initial mass of each model is indicated by the color of the scatter point. We set the outburst timescale to a minimum value of $10^{-2}$ years for models that do not exceed $10^{47}$ erg by $10^{-2}$ years before core collapse.
    }
    \label{fig:outburstevst}
\end{figure}

Figure \ref{fig:outburstevst} shows where each of our models lies on a plot of outburst energy versus outburst timescale. We set a minimum outburst timescale of $10^{-2}$ years, so models which do not accumulate $10^{47}$ erg of energy by $10^{-2}$ years before core collapse are plotted with their integrated wave heat at a time of $10^{-2}$ years. Thus the figure shows a group of models at $t_{\rm burst} = 10^{-2}$ years which have accumulated $E_{\rm burst} < 10^{47}$ erg of energy by this time. 

The majority of models which do reach $E_{\rm burst} > 10^{47}$ erg by $10^{-2}$ years before core collapse are clustered just above the threshold energy and time. Thus we expect most stars with ``outbursts" will lie at the weak end of the distribution, and their outbursts may only become apparent in the final days before explosion. It is important to remember that because of the time it takes acoustic waves to propagate to the surface, there is a substantial delay between the wave generation and the deposition of energy in the envelope. This delay time will be approximately $t_{\rm delay} \sim \! 0.02 \, {\rm year} (M/10 \, M_\odot)^{-1/2} (R/100 R_\odot)^{3/2}$. In red/yellow supergiants with hydrogen-rich envelopes, there is not enough time for the acoustic waves to propagate to the surface, so these stars would likely exhibit no outburst at all.

However, several models may produce earlier and more energetic outbursts. The $11\,  M_{\odot}$ model is a clear outlier; with $t_{\rm burst} = 10$ years, its outburst timescale is orders of magnitude earlier than most models. Its energy scale is also above average, with $E_{\rm burst} = 2.5 \times 10^{47}$ erg, due to the semi-degenerate ignition of Ne as discussed in Section \ref{sec:notablemodels}. The next earliest model, with initial mass $32\,  M_{\odot}$, has a similar energy scale but an outburst timescale of months. 
The $44\,  M_{\odot}$ model has the highest outburst energy, $E_{\rm burst} = 7.2 \times 10^{47}$ erg. 
Within the group of energetic and early outliers, all but two of the models are very massive ($M>30 \, M_\odot$) and experienced a shell merger event that catapulted the accumulated energy of that model up to $> \! 10^{47}$ erg. The exceptions are the $11\,  M_{\odot}$ and $14\,  M_{\odot}$ models (blue dots at $E_{\rm burst} \approx 2\text{--}3 \times 10^{47}$ erg), both of which experience degenerate ignition of O/Ne or Si burning as explained above

Once the wave energy is transmitted to the base of the stellar envelope, the waves will damp as they travel toward the surface as acoustic waves and deposit their energy in the envelope. Waves steepen and thermalize their energy due to both weak shock dissipation and radiative diffusion damping. This process is described in detail for red supergiants in \citet{fuller2017} and hydrogen-poor progenitors in \citet{fuller2018}. They find that wave energy is typically deposited just above the core,
where the wave heat can increase the pressure of the heated region and cause it to expand.  The expansion is approximately hydrostatic if $t_{\rm heat} \gtrsim t_{\rm dyn}$, but if $t_{\rm heat} < t_{\rm dyn}$, it will launch a pressure wave. For red supergiants, \citet{fuller2017} find that during core Ne burning, $t_{\rm heat} < t_{\rm dyn}$ so that wave heating launches a pressure wave into the envelope, which drives a small outflow  ($M < M_{\odot}$). In addition, wave heating from core O burning, where $t_{\rm heat} \gtrsim t_{\rm dyn}$, inflates the envelope and causes an unusual envelope density structure to form. In stripped progenitors, waves deposited just above the core are very near the stellar surface, so the wave energy deposition also drives an outflow \citep{fuller2018}.

\subsection{Implications for supernovae and their progenitors}

As described in Section \ref{sec:results}, our results for the energy transmitted by wave heating are generally lower than that of prior work since we now account for a spectrum of $\ell = 1 \text{--}10$ waves and for non-linear wave dissipation. We therefore find outbursts to be less common among our models than previously expected; in turn, those models that do have outbursts also have lower outburst energies than the findings of prior work. Our models do not yet include the effects of wave heating on the star's structure and luminosity, but our results allow for basic inference of pre-SN outburst properties.

The outburst energies and timescales of our models indicate that outbursts may be most common among low-mass ($M \lesssim 12\, M_{\odot}$) stars and a  fraction of high-mass ($M \gtrsim 30\, M_{\odot}$) stars. Many of these progenitors would exhibit outbursts on a timescale of days to weeks before core collapse, but the outburst timescale varies considerably. As noted in Section \ref{sec:outbursts}, the delay time for acoustic waves to propagate to the surface and produce a potential outburst could be longer than the time to core-collapse in red supergiants. We may also consider outbursts from stripped SN progenitors, which will have similar outburst timescales and energies to models that share the same He/C core masses since the evolution of core burning in stars is well determined by these quantities (Figure \ref{fig:corevsinitial}). Given the negligible delay-time in stripped SN progenitors, outbursts can occur in these stars even for timescales far closer to core collapse.

For low-mass stars, pre-SN outbursts are most likely to occur years or perhaps even decades before core-collapse for the lowest-mass SN progenitors. Such outbursts would be fueled by high wave energies generated by vigorous convection at the onset of semi-degenerate Ne ignition. We have not simulated $M < 11 \, M_\odot$ stars, whose core evolution can be very complex and difficult to model due to off-center ignition of O/Ne and Si; yet since these elements ignite semi-degenerately, wave-driven outbursts may be common in these stars. Indeed, even in the absence of convectively excited gravity waves, degenerate Si ignition in these stars can launch an energetic hydrodynamic pulse that can partially eject the envelope \citep{Woosley2015} weeks or months before core collapse. Convectively excited waves may increase the energy of such outbursts, as well as the parameter space over which they occur.

In general, we expect that more energetic wave heating is capable of producing more massive ejecta, and earlier outburst timescales will propel the CSM to larger radiii where it can affect the supernova light curve and spectrum. For example, the outburst energy of our $11 \, M_\odot$ model is sufficient to eject a thin ($\sim \! 1\, M_\odot$) hydrogen envelope from a yellow supergiant and accelerate it to $\sim \! 100$km/s. In the ten years between that event and core-collapse, this material could expand out to $\sim\! 3 \times 10^{15} \, {\rm cm}$. This CSM mass, radius, and  velocity is similar to that inferred for transformational SNe such as SN2017dio \citep{Kuncarayakti2018}. It is also similar that inferred for SN2014C \citep{Margutti2017}, though the inferred CSM radius in that event of $> 10^{16} \, {\rm cm}$ is larger by an order of magnitude. A wave-driven outburst from a slightly lower mass progenitor, with longer time until core-collapse, could potentially explain that event.

In a hydrogen-rich envelope, the larger binding energy may prevent total envelope disruption, but the envelope's density profile may be altered and a small amount of marginally bound CSM may also be produced \citep{fuller2017}. This type of very confined CSM potentially contributes to early peaks in some type II-P SNe light curves \citep{Moriya2011,Morozova2016,Das2017,Moriya2018,Morozova2020}. The CSM structure depends on the details of the heating history, as slow and steady heating will inflate the star without producing CSM and will not match observations of type II-P SNe \citep{Ouchi2019}. However, our models typically exhibit sharp spikes in the wave heating rate at the onset of nuclear burning phases. This sudden wave heating can likely launch shocks that propagate through the stellar envelope, potentially unbinding material at the surface rather than inflating the entire star \citep{Morozova2020,Leung2020}. More detailed hydrodynamic modeling should be performed to determine the pre-SN stellar/CSM density profile resulting from these outbursts.

In hydrogen-free progenitors, the ejected mass would form a dense wind \citep{fuller2018} of circumstellar He, perhaps similar to that observed via flash ionization in the heavily stripped type Ib/c SNe iPTF 14gqr \citep{De2019} and SN2019dge \citep{Yao2020}. However, SNe with even larger inferred CSM radii ($R_{\rm CSM} \gtrsim 10^{17} \, {\rm cm}$, such as SN2004dk \citep{Mauerhan2018,Pooley2019}, most likely arise from a different mechanism that can operate at longer timescales ($\gtrsim \! 100$ years) before core-collapse.

More massive progenitors with shorter pre-SN outburst time scales are likely to produce a more confined CSM. For instance, the outburst energies of our $36-40 \, M_\odot$ models are sufficient to eject $\sim  \! 10^{-2} \, M_\odot$ from their He cores (assuming they end their lives as $M \! \sim \! 15 \, M_\odot$ Wolf-Rayet stars), but this material can only expand out to $\sim \! 2 \! \times \! 10^{14} \, {\rm cm}$ before core-collapse. It is thus swept up within the first few days after explosion. However, the shock breakout from this extended CSM would produce an extremely fast rise and fall of the optical light curve, and these CSM parameters are very similar to that inferred for the Ic-BL SN2018gep \citep{Ho2019}. Hence, we believe confined wave-driven CSM from massive progenitors ($M_{\rm ZAMS} \gtrsim 30 \, M_\odot$) may provide a compelling explanation for some fast blue optical transients (FBOTs) like SN2018gep.

\subsection{Caveats}

Our results for wave energy transport still involve several uncertainties in the treatment of the relevant physics. Although we are able to identify waves that should experience non-linear wave breaking and have included an approximation of this attenuation in our calculations, it is difficult to quantify how much the amplitude of each wave will be reduced. Due to non-linear coupling between waves of different $\ell$, it is also not clear exactly when non-linear breaking will occur. A determination of a reliable metric for the onset and efficacy of non-linear damping by extending upon work such as  \citet{kumar1996} and \citet{Weinberg2008} should be performed in the future.

The uncertainty of the convectively excited wave spectrum must also be considered. Our work assumes the spectrum of equation \ref{eq:ellpowerspectrum}, but the true spectrum remains a subject of active research (e.g., \citealt{Lecoanet2013,Couston2018,Edelmann2019}). We also assume all waves are excited at the same mass-weighted average value of $\omega_{\rm con}$, very different from the realistic polychromatic spectrum of waves generated by convection. The effect on wave heating rates is unclear. In the case of higher frequency waves, we would have underestimated $f_{\rm esc}$ (Equation \ref{eq:heatfrac}), and vice versa for low frequency waves. We also used the mass-weighted convective luminosity for each convective burning shell in our calculations, but the gradient in convective frequency and luminosity is often quite steep across these regions. Therefore, it is possible that waves are actually excited at different amplitudes and frequencies than we have assumed. Our models utilize standard MLT theory to model convection, but we note that convective instabilities \citep{Arnett2011,Meakin2011,Smith2014c} may change the nature of convection (thereby changing the wave spectrum), potentially driving outbursts even in the absence of non-radial waves.

We have not modeled the effect that wave heating will have on the structure of the star, focusing here on calculating the amount of power and energy transported by waves. Once the waves reach the envelope, they will heat the envelope and modify its structure as discussed in \citet{fuller2017} and \citet{fuller2018}, but this process should not alter our calculations of the wave heating rate because the core evolution is nearly independent of the envelope. Furthermore, the wave energy dissipated within the core (i.e., the waves that do not escape) will not have a strong effect on its evolution, as the total energy from waves that enters the gravity wave cavity is much smaller than the binding energy of that cavity. For example, the wave energy $E_{\rm wave} = \int L_{\rm wave} dt$ excited throughout O/Ne and Si burning phases in the $11\, M_{\odot}$ model is negligible compared to the binding energy $E_{\rm bind}$ of the overlying g-mode cavity, $E_{\rm wave} \sim 10^{-3} E_{\rm bind}$; for our most energetic $44\, M_{\odot}$ model, O/Ne and Si burning phases only inject $E_{\rm wave} \sim 10^{-2} E_{\rm bind}$ into the overlying gravity wave cavity.

In addition, our 1D stellar evolution calculations cannot capture the multi-dimensional effects that come into play for convection during late nuclear burning stages. \citet{Yadav2020} demonstrate that 3D simulations of nuclear burning in the minutes before core-collapse can have much larger convective velocities than in 1D simulations. An increase in the assumed convective velocities could somewhat increase our wave-heating rates, which scale linearly with the RMS convective velocity for a given convective flux. Several other simulations of carbon and oxygen burning on longer time scales (e.g., \citealt{Meakin2006,Meakin2007a,Arnett2009,Arnett2011}) also indicate that 1D models slightly underestimate RMS convective velocities, and they do not properly capture the stochasticity of convection and entrainment that occurs at convective boundaries. Hence, it seems likely that our models marginally underestimate wave flux, but it remains unclear whether other multi-dimensional effects will significantly alter our results.

Finally, we have treated the convective shell mergers in our simulations with some caution, as spontaneous mixing between convective shells can in some cases be a numerical artifact. We performed resolution tests on models that exhibited shell mergers to test whether they were products of poorly resolved shell boundaries. In our tests, the phenomena as described in Section \ref{sec:notablemodels} usually persisted throughout the increases in resolution, although in some cases details such as the timing of the shell merger changed. Thus the shell merger events were not occurring simply due to insufficient resolution in the cores of our models. Interestingly, 3D simulations of late-stage nuclear burning in general produce even more mixing and shell merger events than in 1D models \citep{Yadav2020}. Nevertheless, it remains uncertain whether adjusting different parameters of our models would have a significant effect on the prevalence of the shell mergers; if so, our results would be substantially altered, as the high-mass models with the greatest outburst potential are each linked to a shell merger and its associated large spike in wave power.

\section{Conclusion}

We have modeled wave heating physics in single-star core-collapse SNe progenitors using a suite of MESA stellar evolution models with solar metallicity and ZAMS masses ranging between $11 \text{--} 50 \, M_{\odot}$. As we evolve the stars until core-collapse, we calculate the wave power that is generated by core convection during late-stage nuclear burning, and the fraction of this energy that is transmitted to the stellar envelope. Our calculations improve on prior efforts by accounting for non-linear damping effects and implementing a more realistic wave spectrum. In most cases, $\ell>2$ waves carry a large fraction of wave power but are more strongly trapped in the core than the $\ell=1$ waves considered in prior work. Hence, much of the wave power is dissipated within the core via neutrino damping and non-linear wave breaking. These effects ultimately reduce our wave heating estimates by $\sim \! 1$ order of magnitude compared to previous results.

In our models for typical SN progenitors ($M_{\rm ZAMS} < 30\, M_{\odot}$), waves excited during oxygen/neon burning typically transmit $\sim \! 10^{46\text{--}47}$ erg of energy between $0.1-10$ years before core collapse. Though we have not simulated the response of the stellar envelope, comparison with \citet{fuller2017} and \citet{fuller2018} indicates this level of energy deposition is unlikely to drive a detectable pre-SN outburst in most SN progenitors. 

There are important exceptions, however, especially in the lowest-mass and highest-mass SN progenitors. Many of our high-mass ($M_{\rm ZAMS} \geq 30\, M_{\odot}$) SN progenitors exhibit convective shell mergers that drive intense nuclear burning. Assuming these events are not numerical artifacts, waves in these models transmit more energy ($\sim \! 10^{47\text{--}48}$ erg) to the envelope, but not until $\sim \! 0.01-0.1$ years before core collapse. We speculate that the confined circumstellar medium created by these outbursts in hydrogen-free stars could lead to rapidly rising and fading transients resembling some Fast Blue Optical Transients.

In low-mass SN progenitors ($M_{\rm ZAMS} \lesssim 12 M_{\odot}$), semi-degenerate neon ignition greatly enhances wave heating due to higher wave fluxes and frequencies. This could drive pre-SN outbursts with energies $\sim \! 10^{47}$ erg, on a time scale of 10 years or longer before core-collapse, which could be related to the CSM found in transitional SNe and some type II-P SNe. Future investigations should further examine the interplay of the complex core evolution and wave heating process in these low-mass stars. In subsequent work, we plan to model the hydrodynamic response of the stellar envelope to wave heating, making more informed predictions for the outburst luminosities, ejecta masses, and CSM density structures.

\section*{Acknowledgments}

This work was partially supported by NASA grants HST-AR-15021.001-A and 80NSSC18K1017. JF acknowledges support from an Innovator Grant from The Rose Hills Foundation, and the Sloan Foundation through grant FG-2018-10515.

\bibliography{bib,library2}
\end{document}